\documentclass[12pt,a4paper]{article}
\usepackage{amssymb}
\usepackage{epsf}
\begin{document}

\title{Gravitational waves from a very strong electroweak phase transition}
\author{\large Leonardo Leitao\thanks{%
E-mail address: lleitao@mdp.edu.ar}~  and Ariel M\'{e}gevand\thanks{%
Member of CONICET, Argentina. E-mail address: megevand@mdp.edu.ar} \\[0.5cm]
\normalsize \it IFIMAR (UNMdP-CONICET)\\ \normalsize \it Departamento de
F\'{\i}sica, Facultad de Ciencias Exactas y Naturales, \\ \normalsize \it
UNMdP, De\'{a}n Funes 3350, (7600) Mar del Plata, Argentina }
\date{}
\maketitle

\begin{abstract}
We investigate the production of a stochastic background of gravitational waves
in the electroweak phase transition. We consider  extensions of the Standard
Model which can give very strongly first-order phase transitions, such that the
transition fronts either propagate as detonations or run away. To compute the
bubble wall velocity, we estimate the friction with the plasma and take into
account the hydrodynamics. We track the development of the phase transition up
to the percolation time, and we calculate the gravitational wave spectrum
generated by bubble collisions, magnetohydrodynamic turbulence, and sound
waves. For the kinds of models we consider, we find parameter regions for which
the gravitational waves are potentially observable at the planned space-based
interferometer eLISA. In such cases, the signal from sound waves is generally
dominant, while that from bubble collisions is the least significant of them.
Since the sound waves and turbulence mechanisms are diminished for runaway
walls, the models with the best prospects of detection at eLISA are those which
do not have such solutions. In particular, we find that heavy extra bosons
provide stronger gravitational wave signals than tree-level terms.
\end{abstract}

\section{Introduction}

In a cosmological first-order phase transition a stochastic background of
gravitational waves (GWs) is generated, which could be observed in proposed
gravitational wave detectors in space \cite{maggioreexp,decigo,bbo,lisa,elisa}.
In particular, gravitational radiation which was formed at the scale of the electroweak
phase transition (i.e., at a temperature $T\sim 100\mathrm{GeV}$) has a
characteristic frequency which, after redshifted, would be around the order of
the mHz today. This is close to the sensitivity range of the observatory eLISA
\cite{elisa}, which is scheduled for launch in 2034. This laser interferometer
has a peak frequency in the range of $10^{-3}-10^{-2}\mathrm{Hz}$.

Although in the minimal Standard Model (SM) the electroweak phase transition
is not of first order, there are several extensions of the SM for which this
transition is strongly first-order. The strength of the phase transition can
be characterized by the value of the Higgs field in the broken-symmetry
phase, $\phi_b$. Thus, a phase transition is usually said to be strongly
first-order if this value fulfils the condition $\phi_b/T> 1$, which is
required for a successful electroweak baryogenesis \cite{ckn93}. However,
for the generation of gravitational waves this condition may be not enough.
A first order phase transition occurs via the nucleation and expansion of
bubbles. The collisions of bubble walls and the motions caused in the plasma
produce gravitational waves. For the GW signal to be sizeable, the bubble
wall velocity as well as the energy which is transferred to fluid motions
must be large enough. This generally requires significant supercooling and
large latent heat. Thus, in principle, the stronger the phase transition,
the larger the amplitude of the generated GWs. 
%There are, however, important
%aspects of the dynamics which must be taken into account, such as, e.g., the
%friction of the bubble walls with the plasma.

The phase transition provides three sources of gravitational waves, namely,
bubble collisions \cite{tw90,ktw92,kt93,hk08,cds08}, magnetohydrodynamic
(mhd) turbulence \cite{kkt94,dgn02,gkk07,kgr08,cds09,kk15}, and sound waves
\cite{gwsound}. The corresponding GW signals have been calculated as functions of a few
relevant quantities which can be computed from the dynamics of the phase transition.
These are the temperature $T_*$, the bubble wall velocity $v_w$, the
duration of the phase transition $\beta^{-1}$, the ratio of the released
energy to the energy in radiation, $\alpha$, and the efficiency factors $\kappa$.
The latter give the fraction of the
available energy which fuels a particular generating mechanism, and have been 
determined as a function of other parameters such as $\alpha$ and $v_w$
\cite{kkt94,m08,ekns10,lm11,lm15}.
These results have allowed for several model-independent studies
\cite{n04,gs07,kkgm08,s15,elisasci}, in which the GW spectrum was computed
as a function of the parameters $\alpha$, $\beta$, $v_w$, and $T_*$. 

Specific models have
also been considered, for which the gravitational radiation was computed as
a function of model parameters
\cite{elisasci,amnr02,hk08b,eknq08,kp10,kkm15,hknr15}. Such a parameter
variation is certainly more realistic,
since %for a given model
the quantities $\alpha,\beta,v_w,T_*$ are not independent. However, these
investigations  rely on several approximations.

In particular, a key
quantity for the calculation is the wall velocity. Unfortunately, computing
the wall velocity is a very difficult task, which involves microphysics
\cite{nrfric} and hydrodynamics \cite{gkkm84,eikr92,ms09}. Complete
calculations of $v_w$ only exist in the non-relativistic approximation
\cite{mp95,js01,k15} and in the ultra-relativistic limit \cite{bm09} (see
also \cite{knr14}). Due to these difficulties, the wall velocity is often
%considered as a free parameter or
set by hand in GW computations. %with some criterion;
For instance, it is very common to assume a Jouguet detonation
\cite{amnr02,hk08b,eknq08,kp10,kkm15}, although this is generally incorrect
\cite{l94}. Alternatively, one may choose a ``representative'' value for a
detonation wall (e.g., $v_w=0.95$ \cite{elisasci}). However, the detonation
velocity may vary between the speed of sound ($c_s\approx 0.58$) and the
speed of light \cite{gkkm84,eikr92,ms09}. Another way of avoiding the
calculation of the wall velocity is to consider runaway walls (i.e.,
$v_w\simeq 1$) \cite{elisasci,hknr15}. The condition for a wall to run away
in a given model is relatively easy to determine. However, there are many
models which do not allow for runway walls.

On the other hand, the quantities $\alpha$, $\beta$, and $v_w$ depend on
the temperature. It is usual to compute these quantities at the nucleation
temperature $T_n$ (and to assume $T_*\simeq T_n$), which is defined as that
at which there is one bubble in a Hubble volume. However, it is well known
that this temperature $T_n$ corresponds to the \emph{onset} of nucleation.
The supercooling continues during the development of the phase transition,
and percolation takes place at a smaller temperature $T_p$ (the value $T_*$
actually corresponds to the reheat temperature after percolation). Since GW
generation begins once bubbles begin to collide, this additional
supercooling should be taken into account, especially for the case of very
strong phase transitions.

In this paper, we consider physical variations of model parameters for
concrete extensions of the SM. Our aim is to study very strong electroweak
phase transitions which allow for detonations or runaway walls. The main
difference of our approach with previous works is that we attempt a
realistic computation of the phase transition dynamics. For each model we
estimate the wall velocity and we follow the nucleation and growth of
bubbles until percolation. We performed a similar analysis in our earlier
work \cite{lms12}. Here we extend that analysis by considering models with
stronger phase transitions, and we improve the calculation by taking into
account the saturation of the friction force at high velocities (which in
particular is responsible for the appearance of runaway solutions). For that
aim we use an approximation for the friction developed in
Refs.~\cite{m13,lm15}, which allows us to treat both detonations and runaway
walls.

The paper is organized as follows. In the next section we describe our
treatment of the dynamics of the phase transition. In Sec.~\ref{gwgen} we
write down the formulas for the GW signal produced by the different
mechanisms, and we discuss the computation of the relevant parameters from
the phase transition dynamics. In Sec.~\ref{results} we present our
numerical results. We discuss on the dynamics of such very strong
electroweak phase transitions and the consequences for the generation of
gravitational waves. In Sec.~\ref{conclu} we summarize our conclusions.

\section{The electroweak phase transition \label{pt}}

\subsection{The free energy}

It is well known that in the SM the electroweak phase transition is a smooth
crossover \cite{bp95}, and one needs to go beyond this model to have a
first-order phase transition. For the SM, the tree-level potential is of the
form
\begin{equation}
V_{0}(\phi )=-m^{2}\phi ^{2}+\frac{\lambda }{4}\phi ^{4}+\frac{\lambda }{4}
v^{4}  \label{v0}
\end{equation}
where $\phi $ is the background Higgs field, defined by $\langle H^{0}\rangle
\equiv \phi /\sqrt{2}$. The parameters in Eq.~(\ref{v0}) are related to the
vacuum expectation value and the mass of the Higgs boson by $v=\sqrt{2/\lambda
}m=246\mathrm{GeV}$, $m_{H}=\sqrt{2\lambda v^{2}} =125\mathrm{GeV}$, and we
have chosen the constant term so that the true-vacuum energy density $V_{0}(v)$
vanishes. The zero-temperature effective potential is given, to one-loop order,
by
\begin{equation}
V\left( \phi \right) =V_{0}\left( \phi \right) +V_{1}\left( \phi \right) ,
\end{equation}
where the renormalized zero-temperature correction is given by
\begin{equation}
V_{1}(\phi )=\sum_{i}\frac{\pm g_{i}}{64\pi ^{2}}\,\left[ m_{i}^{4}(\phi
)\left( \log \frac{m_{i}^{2}(\phi )}{m_{i}^{2}(v)}-\frac{3}{2}\right)
+2m_{i}^{2}(\phi )m_{i}^{2}(v)-\frac{m_{i}^{4}(v)}{2}\right] .  \label{v1}
\end{equation}
Here, the sum is over all particle species which couple to the Higgs, $g_{i}$
is the number of degrees of freedom (d.o.f.) of particle species $i$, the upper
sign correspond to bosons, the lower sign to fermions, and $m_{i}$ are the
Higgs-dependent particle masses. In this expression, we have chosen
renormalization conditions that the tree-level values of the minimum, the Higgs
mass, and the true vacuum energy density, are not shifted by radiative
corrections, i.e., $V_{1}(v)=V_{1}^{\prime }(v)=V_{1}^{\prime \prime }(v)=0$ (a
prime indicates a derivative with respect to $\phi $). The relevant species are
those with larger couplings to the Higgs. For the SM, these are the $Z$ and $
W$ bosons, the top quark, and the Higgs and Goldstone bosons. For simplicity we
shall ignore the Higgs sector in the one-loop radiative corrections.

The dynamics of the phase transition is determined by the free energy
density (also called finite-temperature effective potential). To one-loop
order, it is given by
\begin{equation}
\mathcal{F}(\phi ,T)=V\left( \phi \right) +\mathcal{F}_{1}(\phi ,T),
\label{ftot}
\end{equation}
where the finite-temperature corrections are given by \cite{quiros}
\begin{equation}
\mathcal{F}_{1}(\phi ,T)=\sum_{i}(\pm {g_{i}})T\int \frac{d^{3}p}{(2\pi )^{3}
}\log \left( 1\mp e^{-E_{i}/T}\right) ,  \label{f1loop}
\end{equation}
with $E_{i}=\sqrt{p^{2}+m_{i}^{2}(\phi )}$. As is well known, for $ m_{i}/T\ll
1$ we have the expansion
\begin{equation}
\mathcal{F}_{1}(\phi ,T)=-\sum_{i}\frac{g_{i}c_{i}^{\prime }\pi ^{2}T^{4}}{90
}+\sum_{i}\frac{g_{i}c_{i}T^{2}m_{i}^{2}(\phi )}{24}-\sum_{\mathrm{bosons}}\frac{g_{i}
Tm_{i}^{3}(\phi )}{12\pi }+\mathcal{O}(m^{4}),  \label{vpot}
\end{equation}
where $c_{i}=1$ ($1/2$) and $c_{i}^{\prime }=1$ ($7/8$) for bosons
(fermions). For the particles of the SM, the masses are of the form
\begin{equation}
m_{i}(\phi )=h_{i}\phi ,  \label{masses}
\end{equation}
and we shall assume for simplicity this dependence for the extra species as
well.

Bosons with masses of the form (\ref{masses}) give cubic terms proportional to
$-h_{i}^{3}T\phi ^{3}$ in Eq.~(\ref{vpot}), which can make the phase transition
strongly first-order for large $h_{i}$. The contributions from the SM gauge
bosons are not enough to cause a strongly first-order phase transition. For
that reason, it is usual to consider SM extensions with extra bosons (such as
e.g. the MSSM). However, when the resummed daisy diagrams are included, it
turns out that the extra-particle contributions to the thermal cubic term are
suppressed, except for the transverse polarizations of gauge
bosons\footnote{For a model in which the gauge fields play a relevant role, it
should be taken into account the fact that the finite-temperature effective
potential is not gauge-invariant \cite{wpr11}.}. To take into account this
suppression, we will just subtract the corresponding cubic terms. For the
models we will consider, the strength of the phase transition does not rely on
such thermally-induced cubic terms. We are interested in cases in which the
order parameter $\phi /T$ is not necessarily small enough for the validity of
the approximation (\ref{vpot}). Therefore, we will compute the complete
one-loop correction (\ref{f1loop}), but we shall add a term $ \sum
g_{i}Tm_{i}^{3}/(12\pi )$ (where the sum includes all the bosonic d.o.f.,
except for the transverse polarizations of gauge bosons) to take into account
the suppression of the cubic terms.

For the effective potential (\ref{ftot}), the symmetric value $\phi =0$ is the
absolute minimum  at high temperatures. In general, there is a range of
temperatures in which this minimum coexists with a symmetry-breaking minimum
$\phi _{m}(T)$. At low temperatures, $\phi _{m}$ is the absolute minimum, which
takes the value $\phi _{m}=v$ at zero temperature. The two minima are
degenerate at the critical temperature $ T_{c} $, i.e.,
$\mathcal{F}(0,T_{c})=\mathcal{F}(\phi _{m},T_{c})$. The unbroken- and
broken-symmetry phases are characterized by the free energy densities
$\mathcal{F}_{u}(T)=\mathcal{F}(0,T)$ and $\mathcal{F}_{b}(T)= \mathcal{F}(\phi
_{m}(T),T)$, respectively. The pressure in each phase is given by
$p=-\mathcal{F}$, the energy density is given by $\rho =\mathcal{F}
-Td\mathcal{F}/dT$, and the enthalpy density by $w=\rho +p=-Td\mathcal{F}/dT$.
At the critical temperature, the pressures are equal, $
p_{u}(T_{c})=p_{b}(T_{c})$, but the energy density is different and the latent
heat $L$ is defined by the difference $L=\rho _{u}(T_{c})-\rho _{b}(T_{c})$.

\subsection{Phase transition dynamics}

Below the critical temperature, spherical bubbles of the broken-symmetry phase
nucleate with a rate \cite{nucl}
\begin{equation}
\Gamma (T)\simeq A(T)\,e^{-S_{3}\left( T\right) /T},  \label{gamma}
\end{equation}
with $A(T)=\left[ S_{3}(T)/(2\pi T)\right] ^{3/2}T^{4}$, where $S_{3}$ is
the three-dimensional instanton action
\begin{equation}
S_{3}=4\pi \int_{0}^{\infty }r^{2}dr\left[ \frac{1}{2}\left( \frac{d\phi }{dr
}\right) ^{2}+V_{T}\left( \phi (r)\right) \right] ,  \label{s3}
\end{equation}
and
\begin{equation}
V_{T}(\phi )\equiv \mathcal{F}(\phi ,T)-\mathcal{F}(0,T).  \label{potef}
\end{equation}
The configuration of the nucleated bubble is a solution of the equations
\begin{equation}
\frac{d^{2}\phi }{dr^{2}}+\frac{2}{r}\frac{d\phi }{dr}=\frac{dV_{T}}{d\phi },
\quad \frac{d\phi}{dr}(0)=0,\quad  \lim_{r\to\infty}\phi(r)=0.
\label{eqprofile}
\end{equation}
The function $S_3(T)$ diverges at $T=T_c$ and, hence, we have $\Gamma(T_c)=0$.
As $T$ decreases below $T_c$, $S_3$ decreases and $\Gamma$ grows.

The temperature variation is governed by the equation $ dT/dt=-HT$, and the
expansion rate is given by the Friedmann equation, $H=\sqrt{8\pi G\rho /3}$,
where $G$ is Newton's constant. At the beginning of the phase transition the
energy density is given by $\rho =\rho _{u}(T)$, and we have a simple equation
for $T(t)$. On the other hand, when bubbles of the broken-symmetry phase are
already nucleated, we have regions with different equations of state. If the
temperature $T$ were homogeneous, we would have $\rho _{b}(T)<\rho _{u}(T)$.
However, the energy difference between the two phases goes into reheating of
the plasma as well as bulk fluid motions. For the fast phase-transition fronts
we are interested in (namely, detonations or runaway walls), the released
energy is distributed behind the walls, i.e., inside the bubbles. Therefore,
the temperature $T_u$ outside the bubbles is homogeneous. Since bubbles
nucleate in the symmetric phase, the rate $\Gamma$ depends only on this
temperature. In spite of this, in the Friedmann equation we have to take into
account also the energy density in the broken-symmetry phase. Nevertheless,
energy conservation insures that the average energy density remains essentially
unchanged, i.e., the average energy density $\bar\rho_b$ inside the bubbles
must be essentially the same as outside, $\bar\rho_b=\rho _{u}(T_u)$. Thus, we
have
\begin{equation}
\frac{dT_u}{dt}=-\sqrt{\frac{8\pi G\rho _{u}(T_u)}{3}}\,T_u.  \label{dTudt}
\end{equation}
From Eq.~(\ref{dTudt}) we readily obtain $T_u(t)$ and, thus,
$\Gamma(t)=\Gamma(T_u(t))$.

Let us call $ t_{c}$  the time at which the Universe reaches the critical
temperature. Then,  at time $t>t_{c}$, the development of the phase transition
is characterized by the fraction of volume $f_u(t)$ in the unbroken-symmetry
phase or, equivalently, that in the broken-symmetry phase, $ f_{b}=1-{f}_{u}$.
Taking into account bubble overlapping and the fact that bubbles can only
nucleate in the symmetric phase\footnote{The calculation is more involved for
deflagration bubbles, which are preceded by shock fronts which reheat the
plasma. Thus, the nucleation rate is also suppressed in the reheated regions of
the symmetric phase. For a treatment of this case, see \cite{lms12}.}, we have
\cite{gw81}
\begin{equation}
f_{u}(t)=\exp \left[ -\frac{4\pi }{3}\int_{t_{c}}^{t}dt^{\prime }\,\Gamma
(T_{u}^{\prime })\left( \frac{a^{\prime }}{a}\right) ^{3}R_{b}\left(
t^{\prime },t\right) ^{3}\right],  \label{fb}
\end{equation}
where we use the notation $T_u^{\prime }=T_u(t^{\prime })$, $a^{\prime
}=a(t^{\prime })$, $a$ is the scale factor (which is determined by the equation
$\dot{a} /a=H$), and $R_{b}$ is the radius of a bubble which nucleated at time
$t^{\prime }$ and expanded until time $t$.

The initial bubble radius can be obtained from the bubble configuration at the
nucleation time $t^{\prime }$. Nevertheless, it is well known that for a phase
transition at the electroweak scale this initial radius is negligible in
comparison with the final bubble size. The bubble wall velocity at a given time
$t^{\prime \prime }$ between $t^{\prime }$ and $t$ is a function of the
temperature outside the bubble, $v_{w}^{\prime \prime }=v_{w}(T_{u}^{\prime
\prime })$ (see below). Thus, the bubble radius at time $t$ is given by
\begin{equation}
R_{b}(t^{\prime },t)=\int_{t^{\prime }}^{t}v_{w}^{\prime \prime }\frac{a}{
a^{\prime \prime }}dt^{\prime \prime },  \label{radius}
\end{equation}
where $a^{\prime\prime}=a(t^{\prime\prime})$.

\subsection{The wall velocity}

The equation of motion for the wall can be obtained from the field equation
(see e.g., \cite{mp95}). For a wall propagating towards the positive $z$
direction with velocity $v_{w}(t)$ and Lorentz factor
$\gamma_w=1/\sqrt{1-v_w^2}$, we may assume a field profile of the form $\phi
(z,t)=\phi _{0}(( z-z_{w})\gamma _{w} )$, with $\phi _{0}(z)$ a
function\footnote{A specific $\tanh $ ansatz is often used.} which varies
between the minima $ \phi =\phi _{m}$ and $\phi =0$ in a small range (the wall
width) around the wall position $z_w(t)$ \cite{lm15}. Then, the wall equation
can be written as
\begin{equation}
\sigma _{0}\gamma _{w}^{3}\dot{v}_{w}=\int \frac{\partial \mathcal{F}}{
\partial \phi }\,\frac{\partial \phi }{\partial z}\,dz+\int dz\sum_{i}g_{i}
\frac{dm_{i}^{2}}{d\phi }\frac{d\phi }{dz}\int \frac{d^{3}p}{(2\pi
)^{3}2E_{i}}\delta f_{i}\,\,,  \label{eqwall}
\end{equation}
where $\sigma _{0}=\int \left[ \phi _{0}^{\prime }(z)\right] ^{2}dz$ (notice
that all the integrands vanish outside the wall, where $d\phi /dz=0$), and $
\delta f_{i}$ are the deviations from equilibrium of the particles distribution
functions. The left-hand side of Eq.~(\ref{eqwall}) is just the proper
acceleration of the wall times its surface energy density at rest. Hence, the
right hand side is the net force $F_{\mathrm{net}}$ (per unit area) acting on
the wall, and we may write $F_{\mathrm{net}}=F_{\mathrm{dr}}+F_{\mathrm{fr}}$,
where the driving force $F_{\mathrm{dr}}$ is given by the first term in
Eq.~(\ref{eqwall}) and the friction force $F_{\mathrm{fr}}$ is given by the
last term.

The driving force is relatively easy to calculate. Taking into account the fact
that the temperature varies across the wall, this force can be written as (see
e.g. \cite{lm15}) $F_{\mathrm{dr}}=\mathcal{F}_{u}-\mathcal{F}_{b}-\int
({\partial \mathcal{F}}/{\partial T^{2}}){dT^{2}}$, which can be approximated
by
\begin{equation}
F_{\mathrm{dr}}=\mathcal{F}_{u}-\mathcal{F}_{b}-\left\langle
\frac{\partial\mathcal{F}}{\partial T^{2}}\right\rangle \left(T_{u}^{2}-T_{b}^{2}
\right),
\label{decomp}
\end{equation}
where $\langle Q\rangle $ means averaging the values of a given quantity $Q$ on
each side of the wall, $\langle Q\rangle =(Q_{u}+Q_{b})/2$. For the cases of a
detonation or a runaway wall,  the temperature $T_{u}$ outside the bubble is
given by Eq.~(\ref{dTudt}), while the plasma inside the bubble is reheated.
This reheating can affect significantly the driving force, which is sensitive
to the departures of $T_{u}$ and $T_{b}$ from $T_{c}$.

The friction force is much more involved. To compute the deviations $\delta
f_{i}$ it is necessary to consider a set of integro-differential Boltzmann
equations for the particles with strongest couplings to the Higgs. The
collision terms of the Boltzmann equations involve all the interactions of
species $i$ with all other species. A detailed computation is very difficult,
and has been attempted only for a few models \cite{mp95,js01,k15}. Among other
approximations, the deviations from equilibrium are assumed to be small in
order to work to linear order in the perturbations. As a consequence, these
calculations break down if the wall velocity is close to the speed of sound
(where the hydrodynamics becomes very strong) as well as for ultra-relativistic
wall velocities \cite{knr14}. In the non-relativistic limit, the friction force
is of the form \cite{nrfric} $F_{\mathrm{fr}}^{NR}=-\eta _{NR}\,v_{w}$.
Analytic approximations for the friction coefficient $\eta _{NR}$ were derived
in Refs.~\cite{m04,ms10}. For masses of the form (\ref{masses}) we
have\footnote{The friction coefficient $\eta _{NR}$ receives also a
contribution from infrared boson excitations \cite{m00}. However, this
contribution is suppressed for strongly first-order phase transitions
\cite{m04,ms10}, and we shall neglect it in this work.},
\begin{equation}
\eta _{NR}=\sum_{i}\frac{g_{i}h_{i}^{4}}{\Gamma }\int_{0}^{\phi
_{c}}c_{1}(\phi)^{2}{\phi }^{2}\sqrt{2V_{T_{c}}(\phi )}\,d\phi ,  \label{etath}
\end{equation}
where $\Gamma $ is an average interaction rate, $\phi _{c}=\phi _{m}(T_{c})$,
and
\begin{equation}
c_{1}=\frac{1}{T_{u}^{2}}\int \frac{d^{3}p}{(2\pi )^{3}}\,\frac{1}{E}\frac{
e^{E_{i}/T_{u}}}{(e^{E_{i}/T_{u}}\mp 1)^{2}}.
\end{equation}
For the electroweak phase transition, the quantity $\Gamma $ is typically $
\sim 10^{-2}T$, and can be chosen to fit the results of detailed numerical
calculations. Most of the quantities appearing in Eq.~(\ref{etath}) are not
very sensitive to the temperature, and thus we evaluate them at $T=T_{c}$,
which simplifies considerably the calculation. Nevertheless, we compute $c_{1}$
at $T=T_{u}$ since this factor is exponentially suppressed in the case of
strong supercooling ($c_{1}$ is often calculated to lowest order in $m/T$, but
this approximation breaks down for $T\ll T_{c}$). The approximation
$F_{\mathrm{fr}}^{NR}=-\eta _{NR}\,v_{w}$  assumes a fluid at rest, where $v_w$
is the wall velocity relative to this fluid. We actually have a fluid velocity
$v$ which varies across the wall (see below). To take into account its effect,
we may replace $v_{w}$ with the average $\langle v_w-v\rangle$. Thus, we have
\begin{equation}
F_{\mathrm{fr}}^{NR}=-\eta _{NR}(v_{w}-\bar{v}),  \label{Ffrnr}
\end{equation}
where  $\bar{v }=(v_{b}+v_{u})/2$.

For an ultra-relativistic wall, the interactions are too slow in comparison to
the passage of the wall, and it is not necessary to consider Boltzmann
equations \cite{bm09}. In this case, the complete distribution functions $
f_{i}$ can be computed exactly, and the result is
\begin{equation}
F_{\mathrm{net}}^{UR}=V(\phi _{u})-V(\phi _{b})-\sum_{i}\frac{
g_{i}c_{i}h_{i}^{2}}{24}T_{u}^{2}\phi _{b}^{2}.  \label{Fnet}
\end{equation}
If this net force is positive, then the wall is accelerated. Actually, in the
UR limit $\dot{v}_{w}$ vanishes due to the $\gamma _{w}$ factors in
Eq.~(\ref{eqwall}), which means that the wall velocity is almost constant,
since it is very close to the limit $v_{w}=1$. Nevertheless, the gamma factor
increases with time, as well as the kinetic and gradient energy of the Higgs
field in the wall. On the other hand, if the UR force $F_{\mathrm{net}}^{UR}$
is negative, then the wall will not reach the UR regime and, instead, will
reach a terminal velocity $v_{w}<1$, with $F_{\mathrm{net}}=0$.

To compute this terminal velocity, we must solve the equation
\begin{equation}
F_{\mathrm{dr}}+F_{\mathrm{fr}}=0,
\end{equation}
where $F_{\mathrm{dr}}$ is given by Eq.~(\ref{decomp}) and $F_{\mathrm{fr}}$ is
a velocity dependent function. The expressions (\ref{Ffrnr}) and (\ref{Fnet})
assume either the NR or the UR limit. In order to take into account the
possibility of intermediate cases, we shall use a phenomenological
interpolation for the friction, which was introduced in Ref.~\cite{m13}. In the
reference frame of the plasma in front of the wall we have \cite{lm15}
\begin{equation}
F_{\mathrm{fr}}=-\frac{\eta _{NR}\eta _{UR}\,\gamma _{w}\bar{\gamma}(v_{w}-
\bar{v})}{\sqrt{\eta _{UR}^{2}+\eta _{NR}^{2}\,\gamma _{w}^{2}\bar{\gamma}
^{2}(v_{w}-\bar{v})^{2}}},  \label{Ffrfeneta}
\end{equation}
where $\bar{\gamma}=1/\sqrt{1-\bar{v}^{2}}$ and $\eta _{NR},\eta _{UR}$ are
free parameters which can be set to fit the values of the friction in the NR
and the UR limits. Indeed, for small wall velocities Eq.~(\ref{Ffrfeneta})
becomes $F_{\mathrm{fr}}=-\eta _{NR}(v_{w}-\bar{v})$ while for $
v_{w}\rightarrow 1$ we have $F_{\mathrm{fr}}=-\eta _{UR}$. Thus, $\eta _{NR}$
is the non-relativistic friction coefficient discussed above, while the
parameter $\eta _{UR}$ is just given by the UR friction force. Notice that
Eq.~(\ref{Fnet}) does not give the UR friction  force but the \emph{total}
force $F_{\mathrm{net}}^{UR}$. The first two terms in Eq.~(\ref{Fnet}) give the
zero-temperature value of the force, while the last term comprises the
hydrodynamics effects as well as the friction. For a given model, the UR
friction can be obtained by subtracting from Eq.~(\ref{Fnet}) the UR limit of
the driving force (\ref{decomp}). Thus, we have
$\eta_{UR}=-(F_{\mathrm{net}}^{UR}-F_{\mathrm{dr}}^{UR})$. This coefficient
will depend on the temperature and on parameters of the model.

\subsection{Hydrodynamics}

The forces acting on the wall depend on the fluid variables
$T_{u},T_{b},v_{u},v_{b}$. For detonations and runaway walls the fluid in front
of the wall is unperturbed. Thus, $T_u$ is given by Eq.~(\ref{dTudt}), $v_{u}$
vanishes\footnote{In the case of deflagrations the hydrodynamics is more
involved, since a shock front propagates in front of the wall. In that case,
the boundary conditions for the temperature and the fluid velocity must be
imposed beyond the shock front. To determine $T_u$ and $v_u$ in front of the
wall, the fluid equations in the symmetric phase must be solved.}, and we only
need to determine the values of $T_{b}$ and $v_{b}$. We are also interested in
the complete temperature and velocity profiles behind the wall, in order to
compute the energy injected in the plasma. Both the fluid equations inside the
bubble and the relations between the  variables  on each side of the wall can
be obtained from the conservation of the energy-momentum tensor. For
hydrodynamics calculations it is usual to consider a simplification of the
equation of state (EOS), namely, the bag EOS.

For the models we will consider, the particles are massless in the symmetric
phase. Hence, in this phase the finite-temperature correction to the effective
potential is just given by the first term in Eq.~(\ref{vpot}), and the free
energy density is of the form
\begin{equation}
\mathcal{F}_{u}=\varepsilon _{u}-a_{u}T_{u}^{4}/3,  \label{Fmas}
\end{equation}
where $\varepsilon _{u}=V\left( 0\right) $ is the false-vacuum energy density,
and $a_{u}=\sum \pi ^{2}g_{i}c_{i}^{\prime }/30$ is the radiation constant.
Thus, the pressure is given by $p_{u}=-\varepsilon _{u}+a_{u}T_{u}^{4}/3$, the
energy density is given by $\rho _{u}=\varepsilon _{u}+a_{u}T_{u}^{4}$, and the
enthalpy density is given by $ w_{u}=(4/3)a_{u}T_{u}^{4}$. The equation of
state of the broken-symmetry phase is more involved. The bag approximation
assumes that the free energy density is of the form of Eq.~(\ref{Fmas}) also in
this phase, i.e.,
\begin{equation}
\mathcal{F}_{b}=\varepsilon _{b}-a_{b}T_{b}^{4}/3.  \label{Fmen}
\end{equation}
Therefore, we have $p_{b}=-\varepsilon _{b}+a_{b}T_{b}^{4}/3$, $\rho
_{b}=\varepsilon _{b}+a_{b}T_{b}^{4}$, and $w_{b}=(4/3)a_{b}T_{b}^{4}$. For
this model we have only vacuum and radiation components. The speed of sound in
both phases is given by $c_s=1/\sqrt{3}$. The critical temperature is
determined by the relation $\Delta a T_c^4=3\Delta\varepsilon$, where $\Delta
\varepsilon =\varepsilon _{u}-\varepsilon _{b}$ and $\Delta a =a_u-a_b$, and
the latent heat is given by $L=4\Delta \varepsilon$.

For the bag EOS, the hydrodynamics depends essentially on two variables. These
are the ratio of degrees of freedom, $ a_{b}/a_{u}$, and the ratio of the
released vacuum energy density to the radiation energy density,
\begin{equation}
\alpha =\Delta \varepsilon /(a_{u}T_{u}^{4}).  \label{alfa}
\end{equation}
Thus, for instance, the driving force (\ref{decomp}) becomes
\begin{equation}
F_{\mathrm{dr}}=\Delta \varepsilon \left(
1-T_{b}^{2}T_{u}^{2}/T_{c}^{4}\right), \label{fdr}
\end{equation}
and the temperatures $T_{b}$ and $T_{u}$ are related by
$T_{b}^{4}/T_{u}^{4}=(a_{u}/a_{b})(w_{b}/w_{u})$. For a detonation, the ratio
$w_{b}/w_{u}$, as well as the fluid velocity $ v_{b}$, are functions of $\alpha
$ and the wall velocity $v_{w}$. Analytic expressions can be found in
Ref.~\cite{lm15}. For a runaway wall, instead, these quantities  depend on
$\alpha$ and the UR net force $F_{\mathrm{net}}^{UR}$. We have \cite{lm15}
\begin{equation}
\frac{w_{b}}{w_{u}}=1+3(\alpha -\bar{F}),\quad v_{b}=\frac{3(\alpha -\bar{F})
}{2+3(\alpha -\bar{F})},  \label{tmevmerun}
\end{equation}
where
\begin{equation}
\bar{F}\equiv F_{\mathrm{net}}^{UR}/a_{u}T_{u}^{4}.
\end{equation}
The limit $\bar{F}=0$ of Eqs.~(\ref{tmevmerun}) matches the UR limit $v_w=1$ of
the detonation result. For $\bar F>0$, the wall is accelerated and part of the
energy and momentum accumulate in the wall. As a consequence, the reheating and
fluid motions decrease.

For a given model, we can calculate $\bar F$ from Eq.~(\ref{Fnet}). Then, using
Eqs.~(\ref{fdr}-\ref{tmevmerun}), we readily obtain the UR driving force and,
hence, the friction parameter $\eta _{UR}$ \cite{lm15},
\begin{equation}
\frac{\eta _{UR}}{a_uT_{u}^{4}}=\alpha -\bar{F}-\frac{1}{3}\left(1-\frac{a_b}{a_u}\right)
\sqrt{\frac{
1+3(\alpha -\bar{F})}{a_b/a_u}}.  \label{etaur}
\end{equation}
Once this parameter is computed, the value of the wall velocity in the steady
state regime can be obtained from the equation
\begin{equation}
\Delta \varepsilon \left( 1-T_{b}^{2}T_{u}^{2}/T_{c}^{4}\right) -\frac{\eta
_{NR}\eta _{UR}\,\gamma _{w}\bar{\gamma}(v_{w}-\bar{v})}{\sqrt{\eta
_{UR}^{2}+\eta _{NR}^{2}\,\gamma _{w}^{2}\bar{\gamma}^{2}(v_{w}-\bar{v})^{2}}
}=0  \label{eqvw}
\end{equation}
For a detonation, the fluid velocity in front of the wall vanishes and we have
$\bar{v}=v_{b}/2$.

The generation of gravitational waves depends on the kinetic energy of the wall
as well as that in bulk fluid motions. For detonations or runaway walls, the
fluid kinetic energy density, given by $w\,\gamma ^{2}v^{2}$, is concentrated
behind the wall, in a region which moves supersonically. For the bag EOS, it is
usual to define the efficiency factor $\kappa_{\mathrm{fl}} $ as the fraction
of the released vacuum energy which goes into bulk motions of the fluid,
\begin{equation}
\kappa _{\mathrm{fl}}=\frac{E_K}{\Delta\varepsilon V_b}
=\frac{\rho _{K}}{\Delta \varepsilon } ,  \label{kappa}
\end{equation}
where $\rho _{K}$ is the average kinetic energy density of the fluid inside the
bubble. We shall assume that the  bubbles are spherically symmetric and
propagate with constant velocity $v_{w}$. In this case, the volume of the
bubble is given by $V_{b}=4\pi (v_{w}t)^{3}/3$, and we have
\begin{equation}
\rho _{K}=\frac{1}{V_{b}}\int_{c_s t}^{v_{w}t}4\pi r^{2}dr\,w\,\gamma ^{2}v^{2}.
\end{equation}
The fluid profiles depend on the variable $\xi=r/t$ and, hence, $\rho_K$ does
not depend on the time $t$. For spherical bubbles, the functions
$w(\xi),v(\xi)$ must be solved numerically. For steady-state walls,
$\kappa_{\mathrm{fl}} $ depends only on the parameter $ \alpha $ and the wall
velocity $v_{w}$ \cite{ekns10,lm11,lm15}, while for runaway walls
$\kappa_{\mathrm{fl}} $ depends only on $\alpha $ and $\bar{F}$ \cite{lm15}. We
shall use the fits for $\kappa_{\mathrm{fl}} $ provided in Ref.~\cite{lm15}.

In the runaway case, a fraction of the released vacuum energy goes into kinetic
and gradient energy of the wall. In this case we may define an efficiency
factor $\kappa_{w}$, in analogy to $\kappa _{\mathrm{fl}}$, as \cite{lm15}
\begin{equation}
\kappa _{w}=\Delta E_{w}/(\Delta \varepsilon \Delta V_{b}),
\label{kappawalldef}
\end{equation}
where $\Delta E_{w}$ is the energy accumulated in the wall as its volume varies
by $\Delta V_{b}$. This coefficient is given by \cite{lm15}
\begin{equation}
\kappa _{w}=F_{\mathrm{net}}^{UR}/\Delta \varepsilon =\bar{F}/\alpha .
\label{kappawall}
\end{equation}

Since the real EOS of the broken-symmetry phase is not of the form
(\ref{Fmen}), in order to utilize Eqs.~(\ref{alfa}-\ref{kappawall}), there is
some ambiguity in the definition of the bag parameters $ \varepsilon _{b}$ and
$a_{b}$. For instance, the vacuum energy density released at the phase
transition may be evaluated from the zero-temperature effective potential as
$V(0)-V(\phi _{m})$. This may be identified with the bag constant $\Delta
\varepsilon $, although the minimum $\phi _{m}$ is temperature dependent. On
the other hand, for the bag EOS we have the relation $L=4\Delta \varepsilon $.
Hence, we could alternatively define the bag constant as $\Delta \varepsilon
=L/4$, with $L=w_{u}(T_{c})-w_{b}(T_{c})$ computed from the effective potential
(\ref{ftot}), so that the bag approximation gives the correct value of the
latent heat. Notice that for a general model we will have
$w_{u}(T_{c})-w_{b}(T_{c})\neq 4(V(0)-V(\phi _{m}))$, and we cannot fit the two
quantities with the bag EOS. A third definition is proposed in
Refs.~\cite{ekns10,gwsound}, namely, $\varepsilon _{b}=(\rho _{b}-3p_{b})/4$,
which is fulfilled for the bag EOS (and, in our case, for the symmetric phase).
This gives $\Delta \varepsilon =(\Delta \rho -3\Delta p)/4$, with $\Delta \rho
=\rho _{u}-\rho _{b}$ and $\Delta p=p_{u}-p_{b}$. For a general model, the
quantity $\Delta \varepsilon $ so defined depends on the temperature. At
$T=T_{c}$ we have $p_{b}=p_{u}$ and, thus, $\Delta \varepsilon =\Delta \rho
/4=L/4$, while at $T=0$ we have $\rho_{u,b}=-p_{u,b}$ and, thus, $\Delta
\varepsilon =\Delta \rho =V(0)-V(\phi _{m})$. Hence, for a phase transition
with little supercooling (i.e., $T\simeq T_{c}$), defining $\Delta \varepsilon
$ in this way will give a bag EOS with the same latent heat as the real model,
while for a phase transition with strong supercooling (i.e., $T\ll T_{c}$),
this definition of $\Delta \varepsilon $ will give the correct value of the
vacuum energy density difference.

In order to choose one of these definitions of the bag parameter $\Delta
\varepsilon $, we need to consider which of these quantities (the latent heat
or the vacuum energy density) is more relevant for the dynamics. The latent
heat represents the energy that is released by the phase transition fronts at
the critical temperature, while the vacuum energy difference is released at
zero temperature. The latent heat reheats the plasma, which affects
significantly the wall motion if $T$ is close to $T_{c}$  but not for $T\ll
T_{c}$. On the other hand, the difference $ V(0)-V(\phi _{m})$ gives the net
force which drives the wall motion at $ T=0$ and is thus relevant in the case
of large supercooling. Since we shall consider phase transitions with different
amounts of supercooling, it is convenient to adopt the last of the above
definitions for the bag constant, namely,
\begin{equation}
\Delta \varepsilon =(\Delta \rho -3\Delta p)/4,  \label{deltaeps}
\end{equation}
where we shall compute $\Delta \rho =\Delta \mathcal{F}-Td\Delta
\mathcal{F}/dT$ and $\Delta p=-\Delta \mathcal{F}$ from
Eqs.~(\ref{v0}-\ref{f1loop}).

Similarly, there is not a unique definition of the effective radiation constant
$a_{b}$. Having defined $\varepsilon _{b}$, we may define $a_{b}$ by
approximating the remaining part of $\rho _{b}$ by a radiation component, i.e.,
$a_{b}=(\rho _{b}-\varepsilon _{b})/T_{b}^{4}$. On the other hand, for the bag
EOS we have the relation $\Delta aT_{c}^{4}/3=\Delta \varepsilon $. Thus, if we
used the definition $\Delta \varepsilon =L/4$, we might also define $\Delta
a=3L/(4T_{c}^{4})$. Alternatively, we may define the radiation constant from
the enthalpy density \cite{ekns10},
\begin{equation}
a_{b}=3w_{b}/(4T_{b}^{4}).  \label{ab}
\end{equation}
At $T=T_{c}$ these definitions coincide. We shall use the latter, with $ w_{b}
$ computed from (\ref{v0}-\ref{f1loop}), since at a given $T$ this definition,
together with the one we already adopted for $\varepsilon _{b}$, allow to write
the real energy density and pressure in the bag form $p_{b}=-\varepsilon
_{b}+a_{b}T_{b}^{4}/3$, $\rho _{b}=\varepsilon _{b}+a_{b}T_{b}^{4}$ (with
$T$-dependent $\varepsilon_b$ and $a_b$).

\section{Gravitational wave generation in a first-order phase transition}
\label{gwgen}

\subsection{The mechanisms of GW generation}

The spectrum of gravitational radiation is usually expressed in terms of the
energy density per logarithmic frequency $d\rho _{GW}/d\log f$, divided by the
critical energy density today $\rho _{c}=3H_{0}^{2}/(8\pi G)$, where $ H_{0}$
is the Hubble rate today, given by $H_{0}=100\,h\,\mathrm{km\,s}^{-1}
\mathrm{Mpc}^{-1}$, and $h=0.72$. Thus, we have a dimensionless quantity
\begin{equation}
h^2\Omega _{GW}(f)=h^2\frac{1}{\rho _{c}}\frac{d\rho _{GW}}{d\log f}.
\end{equation}
Several mechanisms of gravitational wave  generation have been considered in
the literature, namely, bubble collisions, turbulence, and acoustically
generated GWs.

The source of gravitational waves are the spatial components of the
transverse-traceless part of the stress-energy tensor $T_{ij}$. For a vacuum
phase transition, i.e., the case of a scalar field without the hot plasma, the
relevant piece is $\partial _{i}\phi \partial _{j}\phi $ \cite{ktw92}.
Therefore, the source of GWs is concentrated inside the thin regions where $
\phi $ varies, i.e., the bubble walls. In this case, it was shown that the
system of colliding bubbles can be approximated by a set of overlapping spheres
with infinitely thin interfaces \cite{kt93}. In this ``envelop approximation'',
only the uncollided parts of the spherically symmetric walls are taken into
account in the computation of GWs. Thus, the relevant quantity to compute from
the scalar field is the integral of $(\partial _{r}\phi )^{2}$ for a single
bubble configuration. In the absence of friction, the walls quickly accelerate
to the speed of light, and the kinetic and gradient energies of the bubble wall
become equal. Hence, the relevant quantity is the total energy of the wall,
which, for a vacuum phase transition is given by the total vacuum energy
liberated by the bubble \cite{kt93}, i.e., $E_{w}=4\pi \int drr^{2}(\partial
_{r}\phi )^{2}=\Delta \varepsilon V_{b}$, with $V_{b}=4\pi t^{3}/3$.

For a thermal phase transition, the bubble walls often reach a terminal
velocity due to the friction with the plasma. Generally, this happens in a time
which is very short compared to the duration of the phase transition. As a
consequence, the walls move with constant velocity, and the energy which is
released in the phase transition goes into reheating of the plasma and into
bulk fluid motions. These fluid motions also produce gravitational waves. In
this case, the relevant part of $T_{ij}$ is $w\gamma ^{2}v_{i}v_{j} $. For
spherically symmetric walls, this gives the kinetic energy density $w\gamma
^{2}v^{2}$. The envelop approximation can still be used \cite{kkt94} if the
kinetic energy of the fluid is assumed to be concentrated in a thin region
around the wall. The essential difference is that in this case $E_{w}$ is
negligible and the result depends on $E_{ \mathrm{fl}}=4\pi \int
drr^{2}w\,\gamma ^{2}v^{2}=\kappa _{\mathrm{fl} }\Delta \varepsilon V_{b},$
where the efficiency factor $\kappa_{\mathrm{fl}} $ is defined in
Eq.~(\ref{kappa}), and  $V_{b}=4\pi v_{w}^{3}t^{3}/3$.

In the case of a runaway wall, we have $v_{w}=1$ like in the case of a vacuum
phase transition. For a runaway wall, an important fraction of the released
energy (but not all of it) goes to the bubble wall, i.e., $E_{w}=\kappa
_{w}\Delta \varepsilon V_{b}$, with $\kappa _{w}$ defined in
Eq.~(\ref{kappawalldef}). The rest of the vacuum energy goes to reheating and
fluid motions, and the relevant quantity for the generation of GWs is the
energy in the thin shell around the bubble wall, $E_{w}+E_{\mathrm{fl}}=(\kappa
_{w}+\kappa _{\mathrm{fl}})\Delta \varepsilon V_{b}$.

Thus, the general result of the envelop approximation can be written as a
function of the ratio
\begin{equation}
\frac{E_{w}+E_{\mathrm{fl}}}{E_{\mathrm{tot}}}=\frac{(\kappa _{w}+\kappa _{
\mathrm{fl}})\Delta \varepsilon }{\rho _{\mathrm{tot}}}\equiv \frac{\kappa
\Delta \varepsilon }{\rho _{\mathrm{tot}}},  \label{oms}
\end{equation}
where $\kappa _{w}$ vanishes for $F_{\mathrm{net}}<0$ and is given by
(\ref{kappawall}) for $F_{\mathrm{net}}>0$, and $E_{\mathrm{tot}}=\rho
_{\mathrm{tot}}V_{b}$ is the total energy per bubble. As discussed in the
previous section, due to energy conservation $\rho _{\mathrm{tot}}$ does not
vary significantly during the phase transition, and can be approximated by
$\rho _{\mathrm{tot}}=\rho _{u}=a_{u}T_{u}^{4}+\varepsilon _{u}$. In many
calculations, the vacuum energy density is assumed to vanish in the broken
symmetry phase (which is not necessarily the case at $T\simeq T_{c}$). In that
case we have $\varepsilon _{u}=\Delta \varepsilon $, and we can write $ \rho
_{\mathrm{tot}}=a_{u}T_{u}^{4}(1+\alpha )$, which gives
\begin{equation}
\frac{\kappa \Delta \varepsilon }{\rho _{\mathrm{tot}}}=\frac{\kappa \alpha
}{1+\alpha }.  \label{rhokrho}
\end{equation}
We shall use instead the expression (\ref{oms}), with $\Delta \varepsilon $
computed from Eq.~(\ref{deltaeps}) and $\rho _{\mathrm{tot}}= \rho _{u}$.
Nevertheless, we have checked that the approximation (\ref{rhokrho}) does not
introduce significant differences in general.

The mechanism described by the envelop approximation is called the
\textquotedblleft bubble collision\textquotedblright\ mechanism. In these
calculations, the phase transition is simulated by nucleating spherical bubbles
at random locations, assuming an exponential nucleation rate $\Gamma (t)=\Gamma
_{0}\exp [\beta (t-t_{0})]$ and a constant wall velocity $v_{w}$. To apply the
results of such a calculation to a specific model, $\beta $ can be obtained
from
\begin{equation}
\frac{\beta }{H}=-\frac{T}{\Gamma }\frac{d\Gamma }{dT}\simeq T\frac{
d(S_{3}/T)}{dT},  \label{defbeta}
\end{equation}
with $S_{3}$ given by Eq.~(\ref{s3}). We shall use the results of
Ref.~\cite{hk08} for the spectrum that would be observed today,
\begin{equation}
h^{2}\Omega _{\mathrm{env}}=1.67\times 10^{-5}\left( \frac{\kappa \Delta
\varepsilon }{\rho _{\mathrm{tot}}}\right) ^{\!\!2}\left( \frac{100}{g_{\ast
}}\right) ^{1/3}\left( \frac{H}{\beta }\right) ^{2}\frac{0.11v_{w}^{3}}{
0.42+v_{w}^{2}}\frac{3.8\left( f/f_{\mathrm{env}}\right) ^{2.8}}{1+2.8(f/f_{
\mathrm{env}})^{3.8}},  \label{omcol}
\end{equation}
where $f_{\mathrm{env}}$ is the peak frequency today, which is obtained by
redshifting the peak frequency $f_{\mathrm{env}\ast }$ at the moment of the
phase transition. The peak frequency is determined by the time scale $\beta
^{-1}$. More precisely, it is given by $f_{\mathrm{env}\ast }\simeq
0.62/(1.8-0.1v_{w}+v_{w}^{2})$ $\beta $, which gives
\begin{equation}
f_{\mathrm{env}}\simeq 1.65\times 10^{-5}\,\mathrm{Hz}\left( \frac{g_{\ast }
}{100}\right) ^{1/6}\left( \frac{T}{100\,\mathrm{GeV}}\right) \left( \frac{
0.62}{1.8-0.1v_{w}+v_{w}^{2}}\right) \frac{\beta }{H}.  \label{fpcol}
\end{equation}
Notice that the factors of $H$ in these expressions are introduced in order to
normalize the constant $\beta $, since one can obtain the ratio $\beta /H$ from
the phase transition dynamics. The quantities $g_{\ast }$ and $T$ appear by
assuming that after the phase transition all the vacuum energy is converted
into radiation, and thus  setting $\rho _{\mathrm{tot}}=(\pi ^{2}/30g_{\ast
})T^{4}$ in the Friedmann equation $H=\sqrt{8\pi G\rho _{\mathrm{tot}}/3}$.
Hence, the number of degrees of freedom  $g_{\ast }$ corresponds to the
broken-symmetry phase and is given by
\begin{equation}
g_{\ast }=(30/\pi ^{2})a_{b},
\end{equation}
which can be obtained from Eq.~(\ref{ab}). For strongly first-order phase
transitions there can be significant supercooling, which means that the
temperature  at which bubbles  nucleate can be quite smaller than the critical
temperature $T_{c}\simeq 100\mathrm{GeV}$. There will also be some reheating at
the end of the phase transition, and the final temperature $T_{ \mathrm{reh}}$
after reheating is the one that should be used in Eq.~(\ref{fpcol}). We will
estimate this temperature from $\rho _{\mathrm{tot}}=a_{u}T_{p}^{4}+
\varepsilon _{u}=a_{b}T_{\mathrm{reh}}^{4}$ (neglecting $\varepsilon _{b}$),
where $T_p$ is the percolation temperature. Nevertheless, in the case of strong
supercooling, $T_{\mathrm{reh}}$ will be much closer to $T_{p}$ than to
$T_{c}$.

The fluid motions caused by bubble wall motions may remain long after the phase
transition is completed. This provides more efficient mechanisms of GW
generation. If the fluid motion is turbulent, the eddies can act as a source of
GWs after the bubbles have collided. In particular, since the plasma is fully
ionized, mhd turbulence can source GWs for several Hubble times. Computing the
turbulence spectrum is very difficult. Analytic calculations \cite{cds09}
assuming a characteristic stirring scale $L_{S}$ give a peak frequency
$f_{\mathrm{turb} \ast }\simeq 3.5/L_{S}$. A fit to these results for the GW
spectrum is given in Ref.~\cite{cds10} (see also \cite{bbcd12,elisasci}),
\begin{equation}
h^{2}\Omega _{\mathrm{turb}}=3.35\times 10^{-4}
\left( \frac{\kappa _{\mathrm{fl}}\Delta \varepsilon }{\rho _{\mathrm{tot}}}\right) ^{\!\!3/2}
\left( \frac{100}{g_{\ast }}\right) ^{1/3}\frac{L_{S}H}{2}\frac{(f/f_{\mathrm{turb}})^{3}
}{(1+f/f_{\mathrm{turb}})^{11/3}\left( 1+8\pi f_{\ast }/H\right) },
\label{omturb}
\end{equation}
where the peak frequency today is given by
\begin{equation}
f_{\mathrm{turb}}=2.7\times 10^{-5}\,\mathrm{Hz}\left( \frac{g_{\ast }}{100}
\right) ^{1/6}\left( \frac{T}{100\,\mathrm{GeV}}\right) \left( \frac{2}{
L_{S}H}\right) .  \label{fpturb}
\end{equation}
In the denominator of Eq.~(\ref{omturb}), $f_{\ast }$ is the frequency at the
time of production. In terms of the redshifted frequency $f$ we may write $
f_{\ast }/H=(3.5f)/(L_{S}Hf_{\mathrm{turb}})$. Equations
(\ref{omturb}-\ref{fpturb}) are usually given in terms of the parameters
$v_{w}$ and $\beta $ rather than $L_{S}$. However, we remark that in
\cite{cds09} the spectrum is calculated as a function of the stirring scale
$L_{S}$. This scale is then estimated as $L_{S}=2R_{b}\sim 2v_{w}\beta ^{-1}$,
i.e., the typical bubble radius $R_{b}$ is assumed to be given by the wall
velocity and the characteristic time scale $\beta ^{-1}$. However, as we shall
see, for very strong phase transitions the quantity $\beta $ does not give a
correct estimation of the time scale. Hence, we prefer to consider a different
length scale for the source of turbulence, such as the average distance between
bubbles.

Recently, numerical simulations of the field-fluid system were performed
\cite{gwsound}, finding no indication of fluid turbulence. Nevertheless, it was
found that sound waves are a long lasting source of GWs. A simple effective
potential was considered, which allows to relate readily its parameters to
thermodynamic quantities such as the critical temperature, the latent heat, or
the surface tension. However, the number of nucleated bubbles as well as the
nucleation temperature $T_{n}$, are set by hand. All the bubbles are nucleated
simultaneously and, hence, reach approximately the same size $R_{b}$ at
collision. Thus, a free parameter of the simulation is the average distance
between bubbles,
\begin{equation}
d=n_{b}^{-1/3},  \label{d}
\end{equation}
where $n_{b}$ is the average number density of bubbles. A fit to the resulting
spectrum is given in Ref.~\cite{elisasci} as a function of the length scale
$v_{w}\beta ^{-1}$, although $\beta $ is not actually a free parameter in the
simulations of Ref.~\cite{gwsound}. The dependence on $\beta $ is obtained by
estimating the average distance as $d=(8\pi )^{1/3}v_{w}/\beta $. This relation
results from an analytic estimation of $n_{b}$ for a nucleation rate of the
form $\exp (\beta t)$ \cite{eikr92}, where $n_b$ is evaluated at a time $t$
such that the fraction of volume in the symmetric phase has fallen to
$f_{u}=1/e$. We will instead compute directly the average distance $d$ from
Eq.~(\ref{d}), where the number density of bubbles at time $t$ is given by
\begin{equation}
n_{b}(t)=\int_{t_{c}}^{t}dt^{\prime }\,\Gamma \left( T^{\prime }\right)
\left( \frac{a^{\prime }}{a}\right) ^{3}(1-f_{b}).  \label{intnucl}
\end{equation}
In this equation, the scale factor takes into account the dilution of the
number density $\Gamma' dt^{\prime }$ and the factor of $f_{u}$ takes into
account the fact that bubbles only nucleate in the symmetric phase. Thus we
write the GW results from sound waves as
\begin{equation}
h^{2}\Omega _{sw}=2.65\times 10^{-6}\left( \frac{\kappa _{\mathrm{fl}}\Delta
\varepsilon }{\rho _{\mathrm{tot}}}\right) ^{\!\!2}\left( \frac{100}{g_{\ast
}}\right) ^{1/3}\frac{Hd}{(8\pi )^{1/3}}\frac{7^{7/2}(f/f_{sw})^{3}}{\left[
4+3(f/f_{sw})^{2}\right] ^{7/2}}.  \label{omsw}
\end{equation}
The peak frequency $f_{sw\ast }$ is estimated as $f_{sw\ast }=(2/\sqrt{3}
)(8\pi )^{1/3}/d$. After redshifting, we have
\begin{equation}
f_{sw}=1.9\times 10^{-5}\mathrm{Hz}\left( \frac{g_{\ast }}{100}\right)
^{1/6}\left( \frac{T}{100\,\mathrm{GeV}}\right) \frac{(8\pi )^{1/3}}{Hd}.
\label{fpsw}
\end{equation}

\subsection{Computation of parameters from phase transition dynamics}

In order to calculate all the quantities appearing in
Eqs.~(\ref{omcol}-\ref{fpsw}), we shall compute the evolution of the phase
transition as explained in Sec.~\ref{pt}. We shall  solve numerically the
evolution equations (\ref{dTudt}-\ref{radius}). To compute the nucleation rate,
we will solve Eq.~(\ref{eqprofile}) iteratively by the overshoot-undershoot
method, and then we will integrate Eq.~(\ref{s3}). We will track the evolution
of the transition through the fraction of volume $f_{b}(t)$. To do that, we
will compute the wall velocity from Eq.~(\ref{eqvw}) and then integrate
numerically Eqs.~(\ref{radius}) and (\ref{fb}).

Notice that the temperature and, hence, all the relevant quantities $v_{w}$,
$\beta $, $n_{b}$, etc. are constant parameters in the GW equations
(\ref{omcol}-\ref{fpsw}). This is due to the simple modeling of the nucleation
and growth of bubbles in the corresponding calculations, and is very useful for
applications. However, in order to use these results, we need to choose a
representative moment in the development of the phase transition. The bubble
nucleation becomes appreciable when there is at least one bubble in a Hubble
volume $V_{H}=H^{-3}$. This condition is sometimes used to define the
nucleation temperature. A rough estimation of this temperature is thus given by
the condition $S_{3}(T)/T\lesssim 4\log(T/H)\simeq 142$ \cite{eknq08}. However,
as the phase transition goes on, the temperature will descend
further\footnote{For the case of very slow bubble walls,
there can also be a global reheating during the phase transition. In contrast,
for detonations or runaway walls the reheating is localized
in a thin shell behind the walls. In this case, temperature homogenization will
occur at the end of the phase transition.}. For very strong phase transitions
this may be important.  We thus define the \emph{initial} time $t_{i}$ by the
condition $V_{H}n(t_{i})=1$. At $t=t_{i}$ we have $f_b\simeq 0$, while the
phase transition ends when $f_b=1$. Bubbles will effectively begin to meet and
collide once their density and size have become large enough. We shall consider
that moment to be the percolation time $t_{p}$, which is given by the
condition $f_{b}(t_{p})\simeq 0.3$
\cite{lms12}\footnote{Although we are interested in detonations and runaway walls,
we shall consider some (fast) deflagration cases for comparison.
Since a deflagration wall is preceded by a shock front, in this case we define $t_p$
as the time at which ``shock bubbles'' percolate \cite{lms12}.}.

Thus, for using the results of the envelop approximation, we will evaluate the
parameters $\beta $ and $v_{w}$  at $ t=t_{p}$. For the results on the sound
wave mechanism, we will compute the average distance $d$ by evaluating
Eqs.~(\ref{d}-\ref{intnucl}) at $t=t_{p}$. For the turbulence mechanism, there
is an ambiguity in the determination of the stirring scale $L_{S}$ at
percolation, since there are bubbles of very different sizes. Due to the strong
variation of the nucleation rate, smaller bubbles (i.e., those which were
nucleated more recently) are much more abundant than larger ones. On the other
hand, larger bubbles inject more energy into the turbulence. We shall consider
the same scale as in the sound waves mechanism, i.e., $L_{S}=d$. In the next
section we discuss further on this ambiguity.

\subsection{Sensitivity curves for eLISA}

As mentioned above, a phase transition at the electroweak scale generally
produces a GW stochastic background in the mHz range, which may therefore be
observable at eLISA. The architecture of this space-based observatory is still
under debate and several designs are under investigation. Two key design
parameters which affect the sensitivity are the arm length and the
low-frequency acceleration. The detail about these issues is explained in
Ref.~\cite{Klein:2015hvg}. Sensitivity curves for a stochastic background
(cosmological sources) are different from the ones calculated for isolated
(astrophysical) sources. We will consider sensitivity curves for the designs
discussed in Ref.~\cite{elisasci}. The data sheet can be found in
\cite{PetiteauDataSheet}. These designs are generally denoted by  N2A5M5L6,
N2A1M5L6, N2A2M5L4 and N1A1M2L4. In our graphics we will plot the corresponding
sensitivity curves in red, magenta, blue and green, respectively.

\section{Numerical results \label{results}}

It is well known that one way of achieving a strongly first-order electroweak
phase transition is through the addition of bosons with large couplings $h_{i}$
to the Higgs. As already discussed, the one-loop effective potential contains
cubic terms $\sim h_{i}^{3}T\phi ^{3}$ which strengthen the phase transition by
causing a barrier in the free energy. However, such terms are generally
suppressed by higher-loop corrections. Nevertheless, if the couplings $h_{i}$
are large enough, the terms $\sim h_{i}^{4}\phi ^{4}\log \phi $ in the
zero-temperature effective potential (\ref{v1}) may cause a barrier at zero
temperature and it is well known that the phase transition may become very
strong (see, e.g., \cite{eknq08,ms08}). Additional SM singlet scalars also
allow new tree-level terms in the scalar potential, which can induce a barrier
either at finite temperature or already at zero temperature (even without cubic
terms) \cite{bm09,cv93}. These models may give very strong electroweak phase
transitions. Alternatively, a non-renormalizable term such as $\phi
^{6}/\Lambda ^{2}$ can make the phase transition strongly first-order if the
scale $ \Lambda $ suppressing such operator is in the neighborhood of the
electroweak scale \cite{dgw08}.

One may think that the existence of runaway walls depends directly on the
strength of the phase transition. However, as discussed in Ref.~\cite{bm09}, if
the strength of the transition relies on thermal loops, then the wall will not
run away, no matter how strong the phase transition may be. Essentially, this
is due to the fact that the net force (\ref{Fnet}) coincides with the mean
field approximation for the free energy difference between the two phases,
where the cubic and higher powers of $ \phi $ are dropped in the expansion of
the thermal part (\ref{vpot}). Moreover, in the example of the SM extension
with strongly coupled bosons, in spite of the zero-temperature  terms
$\phi^4\log\phi$,  the existence of runaway walls requires  fine tuning of the
parameters \cite{lms12}. Runaway solutions are more likely in extensions with
tree-level terms.

We wish to investigate the generation of GWs in such strong phase transitions,
and in particular to compare models with and without runaway walls. In order to
analyze these general features and to keep our results as model independent as
possible, we will consider simple models representing the different model
classes.

\subsection{The SM with strongly-coupled extra bosons}

Several extensions of the SM consist of adding scalar singlets which are only
coupled to the Higgs. We have considered a simple example of such a hidden
sector in Ref.~\cite{lms12}, consisting of $g$  bosonic degrees of freedom with
Higgs-dependent masses of the form $ m^{2}=h^{2}\phi ^{2}+\mu ^{2}$. The phase
transition is stronger for higher values of the coupling $h$ and for larger
numbers of degrees of freedom, while it becomes weaker for larger $\mu $. Here,
we shall consider only the case $ \mu =0$. The essential difference of our
present treatment with that of \cite{lms12} (for this particular model) is that
we now take into account the saturation of the friction, which leads to higher
values of the wall velocity.

\begin{figure}[bt]
%\centering
\epsfysize=5.2cm \leavevmode \epsfbox{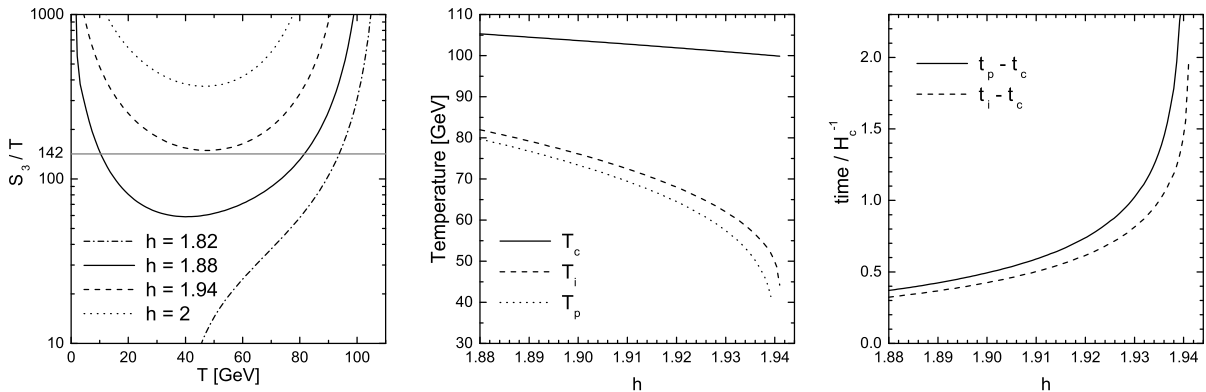}
\caption{Left: thermal instanton
action as a function of temperature,  for $g=2$ and several values of $h$.
Center: the critical temperature $T_c$, the temperature $T_i$ at the onset of nucleation,
and the temperature $T_p$ at percolation. Right: time intervals between the temperature
$T_c$ and the temperature $T_i$ (dashed line) and between $T_c$ and $T_p$ (solid line).}
\label{figtempbos}
\end{figure}

For large enough $h$, the effective potential has a barrier at $T=0$, which is
reflected in the dependence of the nucleation probability on temperature. This
can be appreciated in Fig.~\ref{figtempbos} (left panel), where we consider the
thermal instanton action $S_{3} (T)$ for a few  values of $h$ (for $g=2$).
At $T=T_{c}\sim 100\mathrm{GeV}$, the instanton action diverges, which means
that the nucleation rate vanishes. As $T$ descends below $T_{c}$, $S_{3}$
decreases and the nucleation rate increases. This is because the minima are no
longer degenerate and the barrier between them is smaller. For low values of
$h$, the barrier between minima disappears at a certain temperature $T_{0}$. In
such a case, $S_{3}$ vanishes at $T=T_{0}$ (see the dashed-dotted line).
Therefore, at $T=T_0$ the nucleation rate becomes extremely high, $\Gamma \sim
T^{4}$, and the phase transition will end before the system reaches this
temperature. However, for high enough values of $h$, there is a barrier still
at $T=0$. Due to this barrier, the probability of thermal activation will
decrease again for small temperatures. Indeed, we see that, for higher values
of $h$, $S_{3}/T$ increases again below $T\sim 50\mathrm{GeV}$.

As already mentioned, the nucleation becomes effective for $S_{3}/T\approx
142$. Although this is a rough approximation, the qualitative picture is
correct. We see, for instance, that for $h=1.82$ the phase transition will
occur at $T\approx 90\mathrm{GeV}$ and for $h=1.88$ it will occur at $T\approx
80\mathrm{GeV}$, while for values of $h$ close to $1.94$ it may happen in the
range $40\mathrm{GeV}\lesssim T\lesssim 60\mathrm{GeV}$ and for higher values
of $h$ it will never happen. The amount of supercooling as a function of $h$ is
shown in the central panel of Fig.~\ref{figtempbos}. We see that, indeed, for
$h\simeq 1.94$ the nucleation of bubbles begins at a temperature $T\lesssim
50\mathrm{GeV}$. However, at this point the temperature decreases abruptly as a
function of $h$, and there is a critical value $h_{\max }$ for which the phase
transition occurs at $T=0$. Beyond this maximum value, the system will remain
stuck in the metastable phase, and the phase transition will never end.

In the right panel of Fig.~\ref{figtempbos} we show the time it takes, once the
universe arrives at the critical temperature,  to reach the onset of nucleation
and then to percolate. We see that for $h>h_{\max }$ the universe remains stuck
in the false vacuum and, thus, enters an inflationary era \cite{eknq08,ms08}.
For GW generation we are interested in very strong phase transitions. However,
we are limited by the condition $h<h_{\max }$. Therefore, considering a higher
value of $g$ will not change the situation significantly. Indeed, for $g=12$,
it was found in Ref.~\cite{lms12} that the maximum value of $h$ is smaller,
$h_{\max }\simeq 1.2$. As a consequence, the maximum strength of the phase
transition is not higher than for the case of $g=2$. Moreover, the friction is
larger and the wall velocity smaller. Here we shall consider only the case
$g=2$.

\begin{figure}[bt]
\centering
\epsfysize=6cm \leavevmode \epsfbox{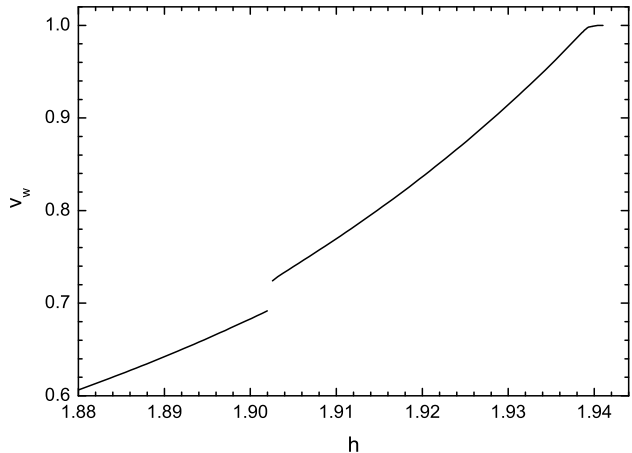}
\caption{The wall velocity for the SM with $g=2$ extra bosons with mass $m=h\phi$.}
\label{figvwbos}
\end{figure}

In Fig.~\ref{figvwbos} we plot the wall velocity for this model, for values of
$h$ which give supersonic  velocities.
In this work we are interested in the detonation and runaway cases.
Nevertheless, we show a small part\footnote{Strongly first-order phase
transitions (i.e., with $\phi/T\gtrsim 1$) exist already for $h\sim 1.2$. In
spite of this, below $h=1.9$ we have deflagration walls. Thus, in comparison,
detonation solutions exist in a  small range. This is due to the large friction
force of these strongly coupled particles.} of the deflagration range for
comparison. For the deflagration case we used the planar wall approximation
(see Ref.~\cite{lms12}). Notice that in this range the deflagrations are
supersonic. The wall velocity has a jump from deflagration to detonation
solutions at $h\simeq 1.9$. In fact, around this value we have coexistence of a
deflagration and a detonation solution. We use the criterion that, in such a
case, the detonation is the one that will be realized in the phase transition
\cite{ms12}. We see that runaway solutions appear very close to the limiting
value $h=h_{\max }$. Therefore, in this model, obtaining runaway walls requires
fine tuning of the parameters.

\begin{figure}[bt]
%\centering
\epsfysize=7cm \leavevmode \epsfbox{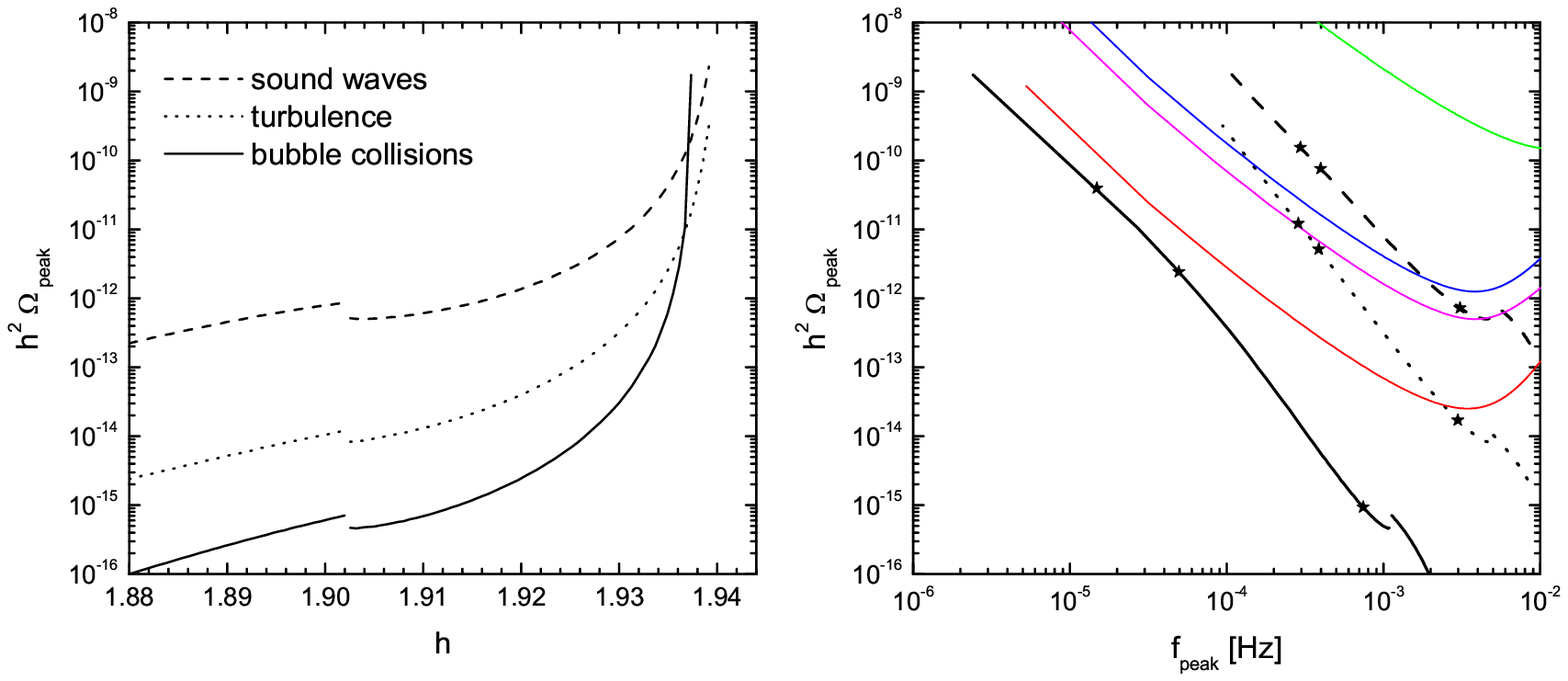}
\caption{Left panel: peak amplitude of the three components of the GW spectrum for the SM with $g=2$ extra
boson d.o.f., as a function of the coupling $h$. Right panel: peak amplitude vs.~peak frequency.
The stars indicate the values of $h$ given in Table~\ref{tabbos}. The colored lines correspond to the
sensitivity curves of eLISA discussed in Sec.~\ref{gwgen}.}
\label{figgwbos}
\end{figure}

In  Fig.~\ref{figgwbos} we consider the GW production for this model.
In the left panel we compare the peak amplitude of the GW spectrum for the
different formation mechanisms\footnote{In Ref.~\cite{lms12} we obtained higher
amplitudes and smaller frequencies for the GWs from turbulence, since we
assumed that the characteristic stirring  length was given by the largest
bubbles.}. Notice that for very strong phase transitions the signal from bubble
collisions surpasses the intensity of the other two, and this happens in the
detonation regime. However, this result should be considered with caution.
Indeed, we find that near $h_{\max }$ the parameter $\beta $ (computed at
$t=t_{p}$) becomes \emph{negative} before reaching the value $h_{\max }$. Thus,
the growth of the bubble-collisions curve is due in part to the blow-up of
$\beta ^{-1}$, while the actual duration of the phase transition is still
finite. This happens because the supercooling is such that the temperature
reaches the minimum of $S_{3}/T$ (we are in the situation of the dashed line in
the left panel of Fig.~\ref{figtempbos}). This indicates that the nucleation
rate is no longer of the form $\exp (\beta t)$, with constant $\beta $ (i.e.,
$S_{3}/T$ should be approximated by a quadratic function of $t$ rather than by
the linear function $\sim \beta t$). Hence, the results
(\ref{omcol}-\ref{fpcol}) from the bubble collision simulation break down. As a
consequence, we cannot compute the signal from bubble collisions beyond a
certain value of the parameter $h$.

In the right panel of Fig.~\ref{figgwbos} we show the peak intensity vs.~the
peak frequency for the same range of values of $h$, together with the
sensitivity curves for the four eLISA configurations mentioned in
Sec.~\ref{gwgen}. We see that, although the intensity of the signal from bubble
collisions grows for very strong phase transitions, its frequency decreases
significantly as well (due to the divergence of $\beta ^{-1}$), departing from
the eLISA sensitivity. On the other hand, the signals from turbulence and sound
waves have similar frequencies, but the one from sound waves dominates.

\begin{figure}[bt]
%\centering
\epsfysize=5cm \leavevmode \epsfbox{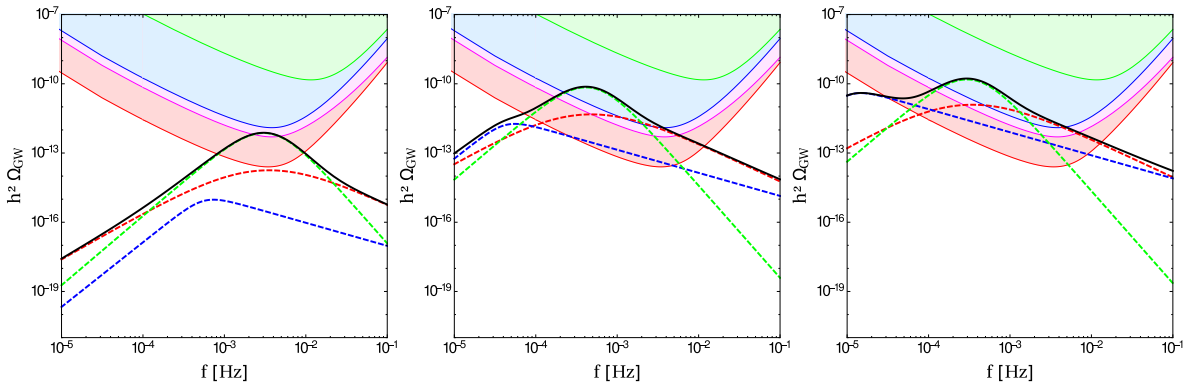}
\caption{The GW spectra for the benchmark points of Table \ref{tabbos}.
The blue dashed line denotes the contribution from bubble collisions, the red dashed line
the contribution from mhd turbulence, the green dashed line the contribution from sound waves,
and the black line the sum of the three signals. The shaded areas represent the regions detectable
by the different eLISA configurations.}
\label{figspecbos}
\end{figure}

The stars on the curves indicate three benchmark points in the range of the
parameter $h$. One of them corresponds to the beginning of the detonation
range, another one corresponds to a fast detonation, and a third one
corresponds to a very strong phase transition, but still with a positive
$\beta$. Some of the properties of the phase transition for these benchmark
points are displayed in Table~\ref{tabbos}. In Fig.~\ref{figspecbos} we plot
the spectra from the different sources and the total spectrum (assuming that
the corresponding contributions to the stochastic GW background combine
linearly \cite{elisasci}).
\begin{table}
\centering
\begin{tabular}{|c|c|c|c|c|c|c|c|} \hline $h$ & $T_p$ [GeV] & $\alpha $ & $\beta /H$ &
$Hd$ & $v_{w}$ & $g_{b}$ & $\phi /T$ \\ \hline 1.913 & 68.1 & 0.057 & 241.5 &
0.013 & 0.789 & 104 & 3.6 \\ \hline 1.936 & 50.3 & 0.187 & 23.8 & 0.075 &
0.967 & 103 & 4.85 \\ \hline 1.937 & 48.3 & 0.219 & 7.4 & 0.098 & 0.977 & 103 &
5.05 \\ \hline
\end{tabular}
\caption{Some characteristics of the electroweak phase transition (the
temperature at percolation, $T_p$, the parameters $\alpha$ and $\beta$, the
average distance $d$ between centers of nucleation, the wall velocity, the
number of degrees of freedom in the broken-symmetry phase at the end of the
phase transition, and the order parameter $\phi/T$ at $T=T_p$)  for a few
benchmark values of the coupling $h$ of the extra bosons with the Higgs.}
 \label{tabbos}
\end{table}

\subsection{The SM with tree-level cubic terms}

In order to investigate the possible effects of tree-level terms, we shall add
a cubic term $\propto \phi ^{3}$ to the tree-level potential. This extension of
the SM must be regarded as a toy model, since such a cubic term cannot be
constructed with the Higgs doublet. In this toy model, the field $ \phi $ may
represent a trajectory in the space of two fields \cite{clw13}. Adding such a
cubic term alone to the potential (\ref{v0}) shifts the tree-level values of
the minimum, the Higgs mass, and the true vacuum energy density. We thus add
the terms
\begin{equation}
-A\phi ^{3}+\frac{3}{4}Av\phi ^{2}+\frac{3}{8}\frac{A}{v}\phi ^{4}-\frac{1}{8
}Av^{3},
\end{equation}
so that  we have $V(v)=0$, $V'(v)=0$, and $V''(v)=2\lambda v^2$. Since the
strength of the phase transition will be dominated by this tree-level
modification, we do not need to consider strongly coupled extra particles.

\begin{figure}[bt]
\centering
\epsfysize=6cm \leavevmode \epsfbox{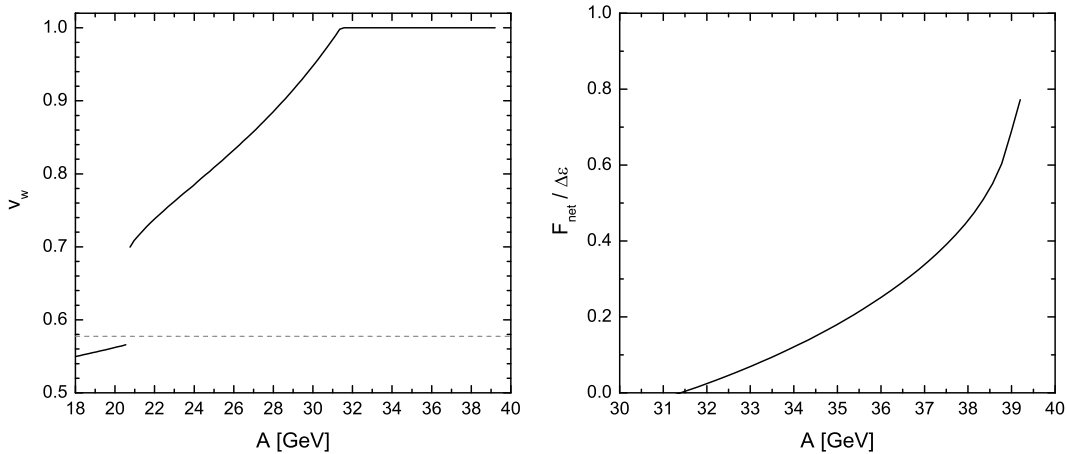}
\caption{Wall velocity (left panel) and net force (right panel) for the model with a cubic
term, as a function of the parameter $A$.}
\label{figvwcub}
\end{figure}

The wall velocity for this model was previously considered in Ref.~\cite{ms10},
obtaining detonations for  $A\gtrsim 20\mathrm{GeV}$, with wall velocities
$v_{w}\lesssim 0.75$. However, the saturation of the friction was not taken
into account. In Fig.~\ref{figvwcub} we plot $v_{w}$ as a function of $A$ (left
panel).
For $A\lesssim 21\mathrm{GeV}$ we have subsonic deflagrations, for
$21\mathrm{GeV}\lesssim A\lesssim 32\mathrm{GeV}$ we have detonations, and for
$A\gtrsim 32\mathrm{GeV}$ we have runaway walls. For $A\simeq 39\mathrm{GeV}$,
the phase transition becomes too strong, like in the previous model. The right
panel of Fig.~\ref{figvwcub} shows the net force normalized to the vacuum
energy density, which also gives the efficiency factor $\kappa _{w}$.

\begin{figure}[bt]
\centering
\epsfysize=7cm \leavevmode \epsfbox{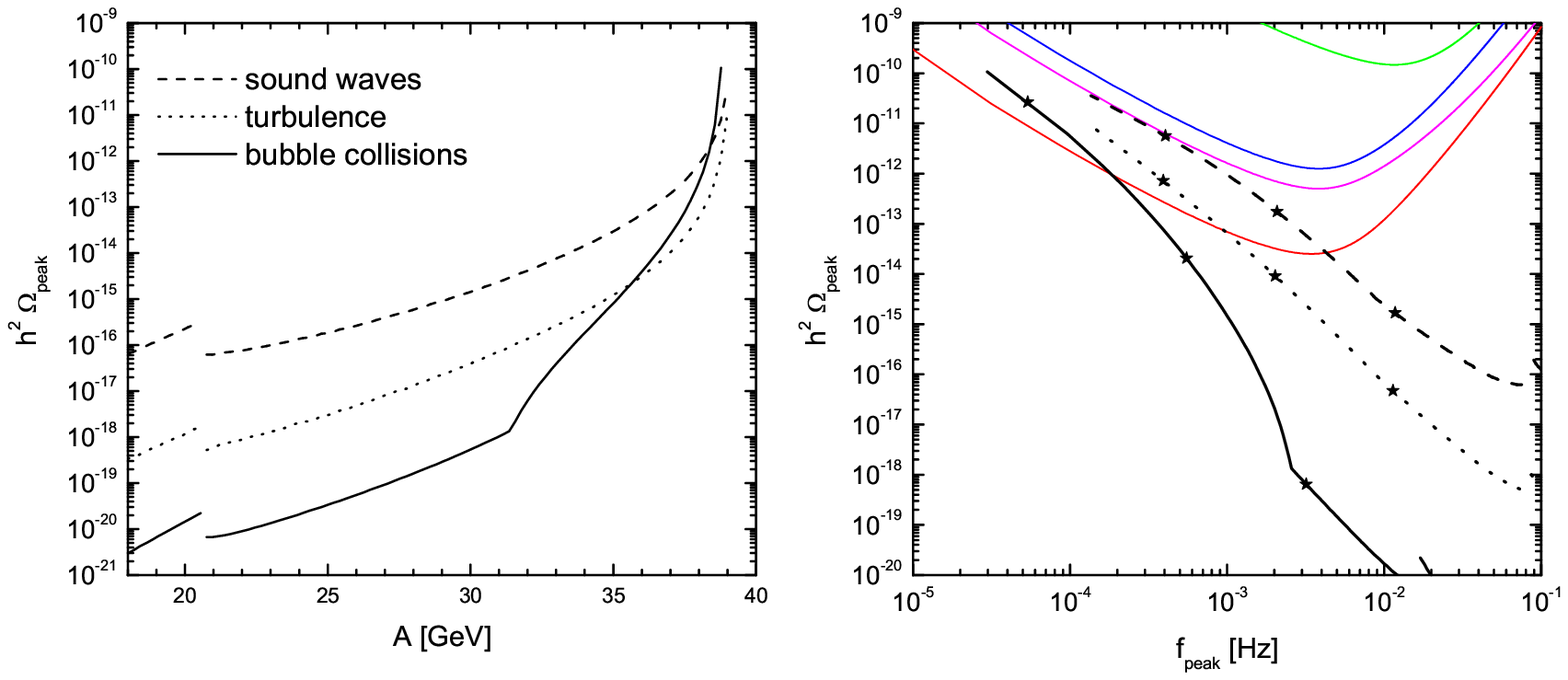}
\caption{As in Fig.~\ref{figgwbos}, but for the model with a cubic term $-A\phi^3$.
The stars indicate benchmark points described in Table \ref{tabcub}.}
\label{figgwcub}
\end{figure}

In the left panel of Fig.~\ref{figgwcub} we plot the peak amplitude of the
spectrum as a function of the parameter $A$.
We see that for detonations the signal from the envelop approximation is
smaller than those from turbulence and sound waves, but it begins to grow as
soon as runaway solutions appear, and gradually approaches the intensity of the
other two signals. In the right panel of Fig.~\ref{figgwcub} we plot the peak
amplitude vs.~the peak frequency for the three signals. Although the amplitude
$\Omega_{\mathrm{env}}$ blows up as the phase transition becomes very strong,
the frequency $f_{\mathrm{env}}$ decreases due to the sudden increase of $\beta
^{-1}$, departing from the peak sensitivity of eLISA. The same happened in the
previous model. Nevertheless, in this case the signal from bubble collisions is
stronger and has a possibility of being observed by the most sensitive eLISA
configuration. This is due to the appearance of runaway solutions. For the same
reason, the signals from turbulence and sound waves do not reach the
intensities they achieved in the previous model.

\begin{figure}[bt]
%\centering
\epsfysize=5cm \leavevmode \epsfbox{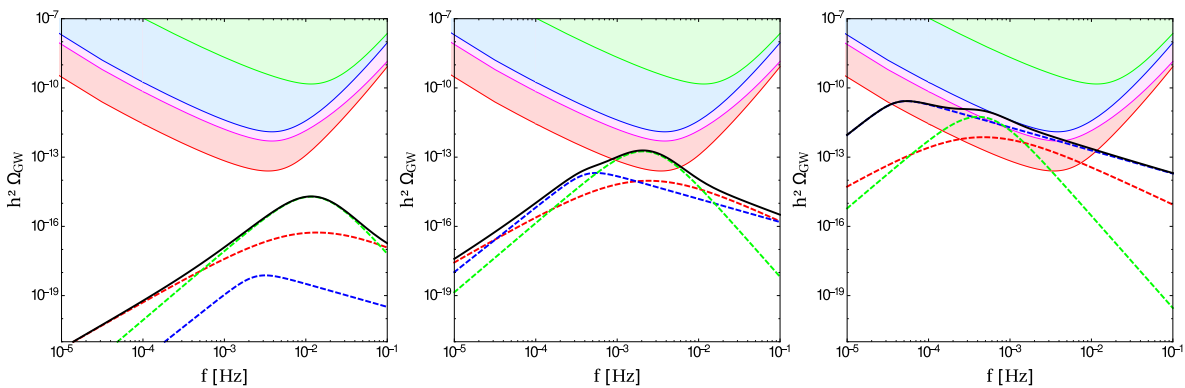}
\caption{As in Fig.~\ref{figspecbos}, but for the benchmark points of Table \ref{tabcub} for
the model with a cubic tree-level term.}
\label{figspeccub}
\end{figure}

We consider again some benchmark points on these curves; one of them in the
detonation range and the other two in the runaway range. Some characteristics
of the phase transition for these points are shown in Table~\ref{tabcub}, and
the corresponding spectra are plotted in Fig.~\ref{figspeccub}.
\begin{table}
\centering
\begin{tabular}{|c|c|c|c|c|c|c|c|}
\hline
$A$ [GeV] & $T_p$ [GeV] & $\alpha $ & $\beta /H$ & $Hd$ & $v_{w}$ & $g_{b}$ & $\phi /T$ \\
\hline
30.3 & 90.3 & 0.021 & 886.6 & 0.004 & 0.958 & 103 & 2.5 \\ \hline
36.9 & 65.3 & 0.079 & 214.9 & 0.018 & 1 & 100 & 3.6 \\ \hline
38.7 & 49.8 & 0.237 & 26.4 & 0.074 & 1 & 96.5 & 4.9 \\ \hline
\end{tabular}
\caption{Some characteristics of the electroweak phase transition (as in Table
\ref{tabbos}) for three benchmark points of the SM with a term $-A\phi^3$.}
 \label{tabcub}
\end{table}

\subsection{The SM with higher dimensional operators}

We shall now consider an effective theory with a sextic term of the form
\begin{equation}
\frac{(\phi ^{2}-v^{2})^{3}}{8\Lambda ^{2}},
\end{equation}
which does not shift the tree-level values of the minimum, Higgs mass, and
vacuum energy density. For $\Lambda \lesssim 840\mathrm{GeV}$ the quartic term
of the potential becomes negative, which is allowed by the presence of the
sextic term. For $\Lambda \lesssim 600\mathrm{GeV}$ the quadratic term of the
zero-temperature potential becomes positive and we have a barrier at zero
temperature. We shall not discuss here the possible tension of this kind of
model with LHC bounds, and we shall only restrict the value of the cutoff
$\Lambda $ by the requirement that the phase transition completes in a few
Hubble times, like in the previous cases. This allows values of the cutoff as
low as $\Lambda \simeq 550\mathrm{GeV}$.

\begin{figure}[bt]
\centering
\epsfysize=6cm \leavevmode \epsfbox{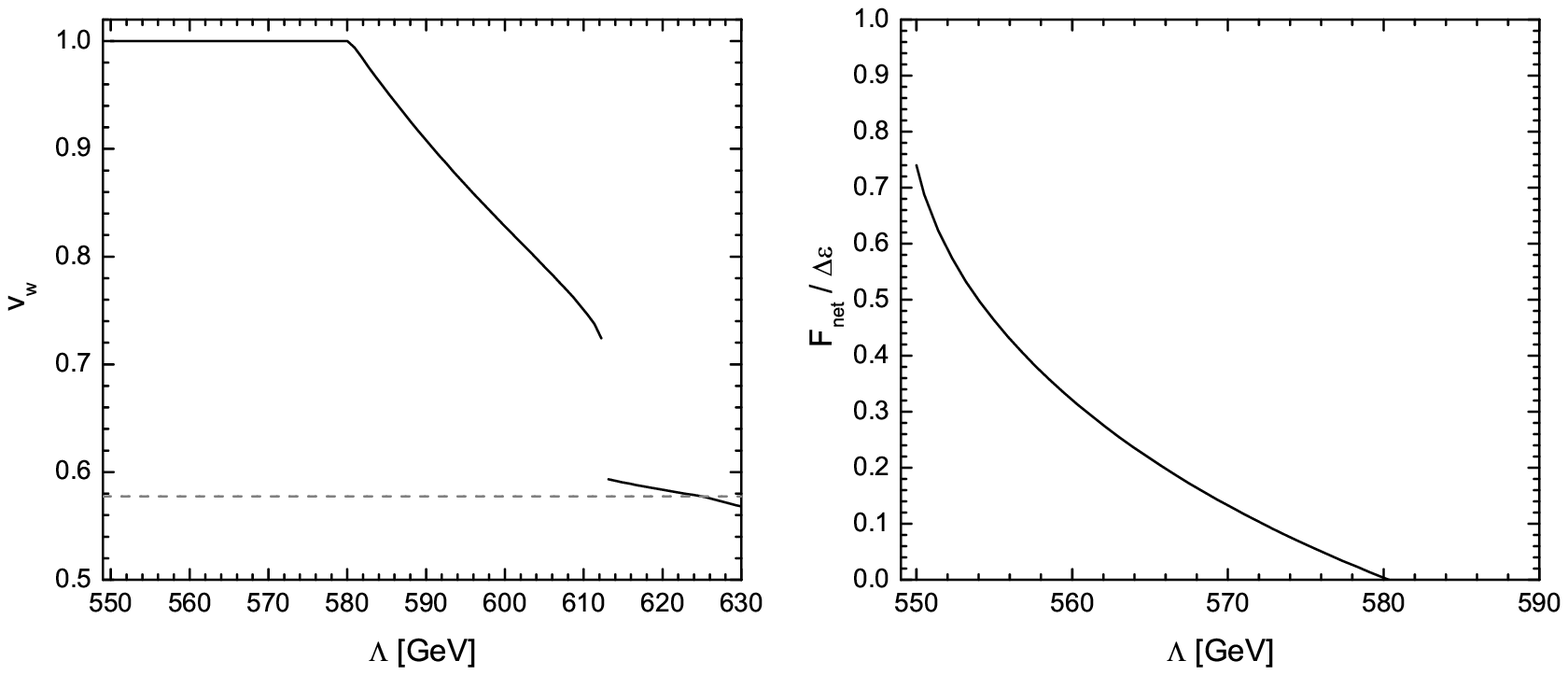}
\caption{The wall velocity (left panel) and net force (right panel) for the SM with
a sextic operator, as functions of the cutoff $\Lambda$.}
\label{figvwsext}
\end{figure}

We show the wall velocity in the left panel of Fig.~\ref{figvwsext} and the net
force in the right panel. There is a detonation range\footnote{This range is
relatively small in comparison with the deflagration range. Indeed, for
$\Lambda\lesssim 840\mathrm{GeV}$ we already have $\phi/T>1$, and we have
deflagrations from this value to $610\mathrm{GeV}$.} for
$580\mathrm{GeV}\lesssim \Lambda\lesssim 610 \mathrm{GeV}$, while below
$\Lambda\approx 580\mathrm{GeV}$ we have runaway walls.

\begin{figure}[bt]
\centering
\epsfysize=7cm \leavevmode \epsfbox{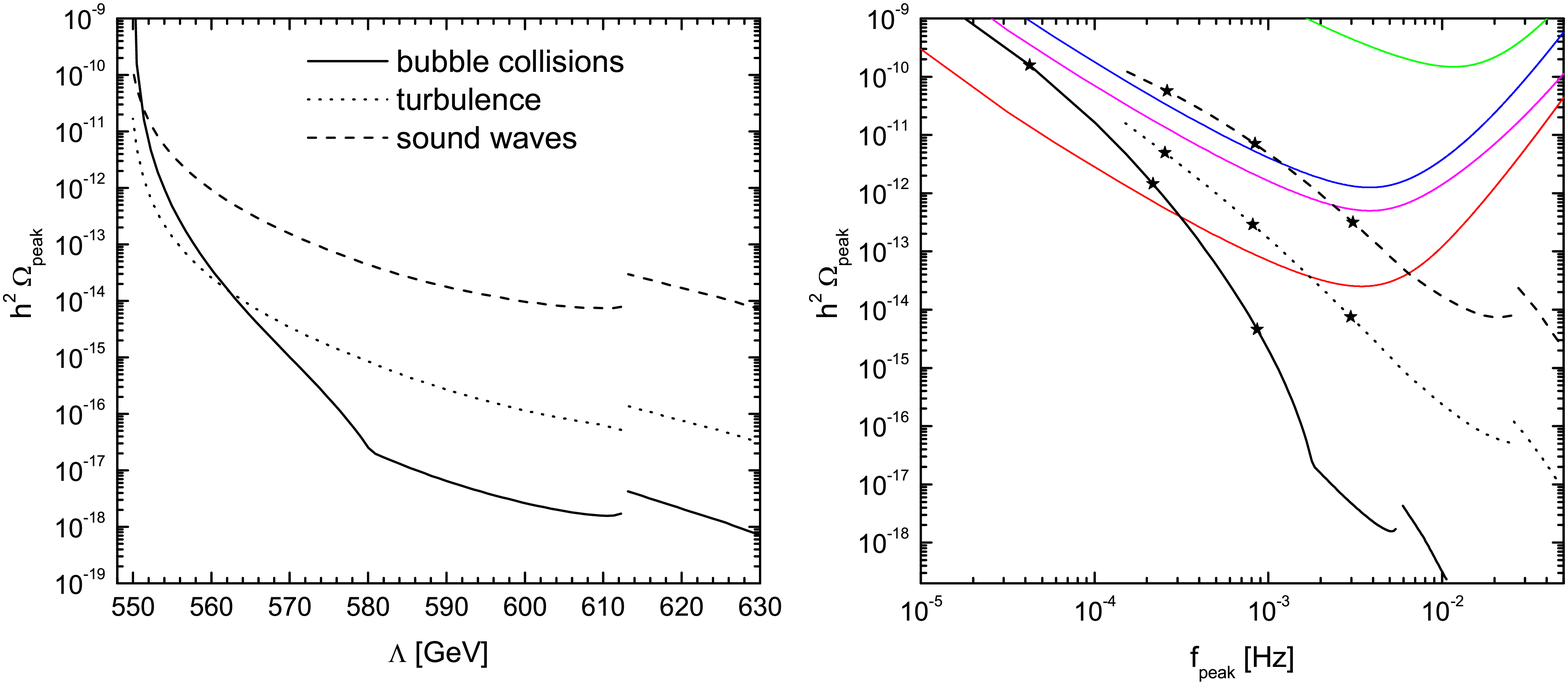}
\caption{As in Fig.~\ref{figgwbos}, but for the model with a sextic term $\phi^6/\Lambda$.
The stars indicate benchmark points described in Table \ref{tabsext}.}
\label{figgwsext}
\end{figure}

\begin{figure}[bt]
\centering
\epsfysize=5cm \leavevmode \epsfbox{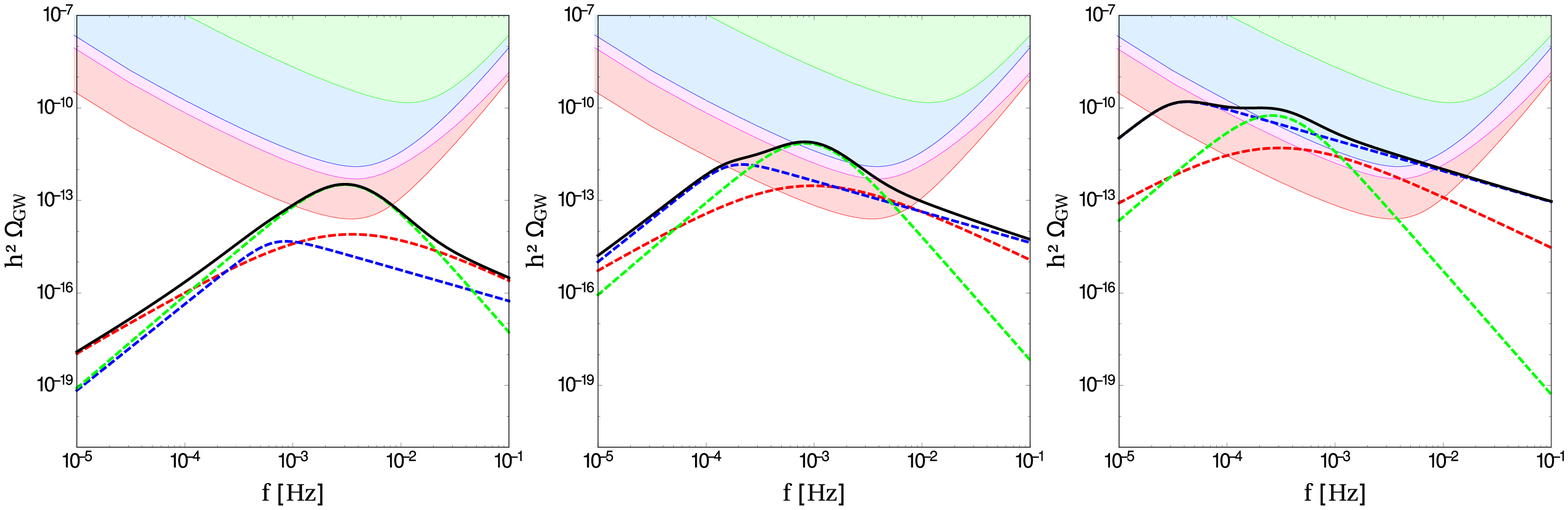}
\caption{As in Fig.~\ref{figspecbos}, but for the benchmark points of Table \ref{tabsext} for
the model with a sextic term.}
\label{figspecsext}
\end{figure}

In Fig.~\ref{figgwsext} we show the GW signals. As in the previous case, the
existence of runaway solutions enhances the signal from bubble collisions,
which gets close to the other signals for very strong phase transitions. On the
other hand, all the intensities are higher than for the previous model.
We have chosen three benchmark points in the runaway region. The corresponding
values of $\Lambda$ and the properties of the phase transition are given in
Table \ref{tabsext}, and the spectra are plotted in Fig.~\ref{figspecsext}.
\begin{table}
\centering
\begin{tabular}{|c|c|c|c|c|c|c|c|}
\hline
$\Lambda$ [GeV] & $T_{p}$ [GeV] & $\alpha $ & $\beta /H$ & $Hd$ & $v_{w}$ & $g_{b}$ & $%
\phi /T$ \\ \hline
565.5 & 57.5 & 0.09 & 379 & 0.01 & 1 & 98.5 & 4.2 \\ \hline
553.5 & 43 & 0.28 & 122 & 0.03 & 1 & 94.5 & 5.7 \\ \hline
550.5 & 34.5 & 0.68 & 27.6 & 0.086 & 1 & 91 & 7.1 \\ \hline
\end{tabular}
\caption{Some characteristics of the phase transition (as in Table
\ref{tabbos}) for three benchmark points of the model with a dimension six
operator with a low cutoff $\Lambda$.} \label{tabsext}
\end{table}

\subsection{Discussion}

For all the  extensions of the SM considered above, there is a range of the
new-physics parameters $h$, $A$, and $\Lambda$ for which the phase transition
has detonations or runaway walls. We observe that models with both kinds of
solutions may give a strong signal of gravitational waves. However, as
expected, the existence of runaway solutions favors the signal from bubble
collisions, while the existence of steady-state walls favors the signals from
turbulence and sound waves. Nevertheless,  even in the runaway range, the
intensity of the signals from fluid motions grows with the strength of the
transition. In particular, the signal from sound waves generally dominates in
the sensitivity range of eLISA. Moreover, it seems that only for phase
transitions with runaway walls the signal from bubble collisions has a chance
of being observable.

\begin{figure}[bt]
\centering
\epsfysize=6cm \leavevmode \epsfbox{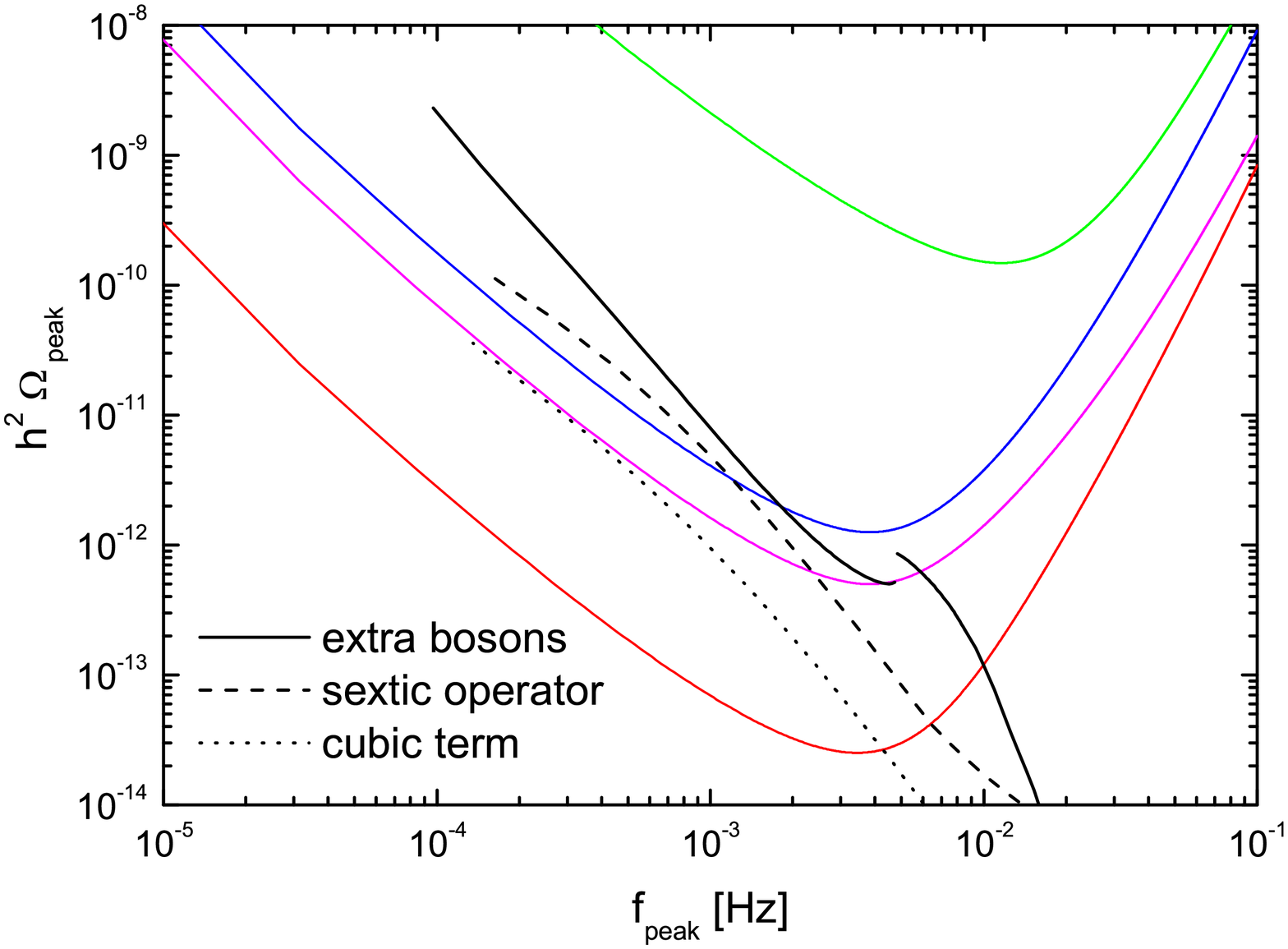}
\caption{The peak amplitude vs.~the peak frequency corresponding to the sound waves
mechanism,
for the three extensions of the SM considered in this work.}
\label{figtodas}
\end{figure}

Since the dominant mechanism is suppressed for runaway walls, the main effect
of the latter is to weaken the  GW signal. As a consequence, the SM extension
with extra bosons gives the strongest signal. In Fig.~\ref{figtodas} we compare
the results from the sound waves mechanism for the different models.

The case of strongly coupled extra bosons clearly deserves further
investigation. Here, we have only considered a mass of the simple form
$m(\phi)=h\phi$ and the particular case of $g=2$ d.o.f. For a more general
mass of the form $m^2=h^2\phi^2+\mu^2$, the phase transition is generally
weaker. On the other hand, for  a larger number of degrees of freedom, the
phase transition is stronger but the friction is larger, which prevents
higher wall velocities. In particular, for $g=12$ we obtain deflagrations in
the whole range of $h$ (up to $h_{\max}$). Nevertheless, as can be seen in
Fig.~\ref{figtodas}, for this kind of model deflagrations may generate an
important GW signal. As already mentioned, deflagrations reheat the plasma
outside the bubbles. This makes the computation of the wall velocity more
difficult, and also complicates the treatment of bubble
nucleation\footnote{To compute the wall velocity in this case we  used the
planar wall approximation and we stopped the evolution of the phase
transition at the percolation of shock bubbles (see \cite{lms12} for
details).}. Therefore, the general treatment of a phase transition mediated
by deflagrations should be improved before embarking in a complete
exploration
of the parameter space. % for the present model.
We shall address this problem
elsewhere.

We remark that  the computation of gravitational waves from phase transitions
is generally affected by ambiguities in the determination of some quantities.
One important source of uncertainty in the cases of turbulence and  sound waves
is the assumption of a single length scale (either by assuming a single
stirring scale or by nucleating all bubbles simultaneously). In a real phase
transition there are different length scales (the sizes of bubbles), and we had
to choose a characteristic size in order to use those results. In the case of
bubble collisions, as already discussed, the approximation
$\Gamma\propto\exp(\beta t)$ breaks down for very strong supercooling. This
introduces an uncertainty in the characteristic time scale $\beta^{-1}$.

\begin{figure}[bt]
\centering
\epsfysize=6cm \leavevmode \epsfbox{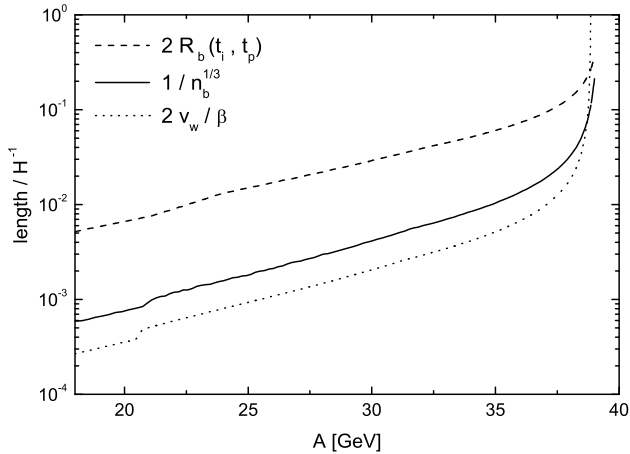}
\caption{Different length scales present in the phase transition,
for the extension of the SM with a tree level term.}
\label{figlength}
\end{figure}

In order to assess the impact of these uncertainties, we compare in
Fig.~\ref{figlength} three different  scales which are often used in the
literature. Besides the average bubble distance $d=n_b^{-1/3}$ used in this
work, we consider the estimation $2v_w\beta^{-1}$ for the bubble diameter, and
the diameter $2R_{b}(t_{i},t_{p})$ of the ``largest bubbles'', i.e., those
which expanded since the onset of nucleation $t_{i}$ until the percolation time
$t_p$. The plot corresponds to the case of the SM with a cubic term; a similar
plot for the case of extra bosons can be found in Ref.~\cite{lms12}.
We see that the average bubble separation is generally an order of magnitude
smaller than the scale of the largest bubbles, and the scale $v_w\beta^{-1}$ is
even smaller. As shown in Ref.~\cite{lms12}, we  have the relation
$R_b(t_i,t_p)\sim 3\log(\beta/H)\,v_w\beta^{-1}$. Had we used the scale $R_b$
instead of $d$, we would have obtained GWs of quite higher intensity but with
smaller frequencies. On the other hand, for very strong phase transitions the
distance $d$ approaches $R_b$, while $v_w\beta^{-1}$ diverges, as already
discussed.

\section{Conclusions} \label{conclu}

We have discussed the generation of gravitational waves in very strong
electroweak phase transitions, such that fast detonations or runaway walls
are possible. For that aim we have studied extensions of the SM with extra
bosons, with tree-level terms, and with non-renormalizable operators. By
considering simple examples of these three classes of models, we focused on
the dynamics of the phase transition. Specifically, we have considered model
parameters which give phase transitions with considerable amounts of
supercooling. We have explored the model parameters up to the limit in which
the universe remains stuck in the metastable vacuum and the phase transition
does not come to an end.

In contrast to earlier studies, we aimed to consider the dynamics of the
phase transition in the most realistic manner possible. In the first place,
we pointed out the uncertainties and ambiguities in the calculations of the
quantities $\alpha$, $\beta$, and other parameters which are relevant for GW
generation, and we discussed the best way to estimate these parameters.
In the second place, we computed the development of the phase transition
from the onset of nucleation to percolation. The calculation included the
evaluation of model-dependent friction parameters which determine the wall
velocity. In this regard, we would like to comment that in recent papers the
computation of the wall velocity was avoided by either considering only
models which give runaway walls \cite{hknr15}, fixing by hand the value of
the detonation velocity \cite{elisasci}, or assuming a Jouguet detonation
\cite{kkm15}.

We have compared the gravitational wave spectra generated via the three
known mechanisms, namely, bubble collisions, fluid turbulence and sound
waves, and we have discussed the possibility of observing these signals at
the eLISA interferometer. For runaway walls, part of the energy goes to the
fluid and part goes to the bubble wall. In comparison with the stationary
motion, this strengthens the signal from bubble collisions and weakens the
signals from turbulence and sound waves. In spite of this, we have found
that the latter two are stronger in most of the parameter regions. This is
because the fluid motions are long-lasting sources of GWs. In any case, the
signal from bubble collisions increases significantly for very strong phase
transitions. This is in part due to the presence of runaway walls, but also
due to the vanishing of the parameter $\beta$, and deserves further
investigation. Nevertheless, in the extreme cases in which the amplitude
from bubble collisions surpasses those from the other sources, its peak
frequency decreases significantly as well. As a consequence, this signal is
never dominant in the region which is detectable by eLISA.

Our main conclusion is that, since the GW signals  from sound waves and
turbulence are generally the dominant ones, detonations are preferable over
runaway walls for phase transitions of similar strength. As a consequence,
among the models we considered, the SM extension with strongly-coupled extra
bosons has the best prospects for being detected at eLISA.

\section*{Acknowledgements}

This work was supported by Universidad Nacional de Mar del Plata, Argentina,
grant EXA699/14, and by FONCyT grant PICT 2013 No.~2786. The work of L.L. was
supported by a CONICET fellowship.

\end{document}